\documentclass{article}
\usepackage{iftex}
\usepackage[utf8]{inputenc}
\ifPDFTeX
    \usepackage[T1]{fontenc}
    \usepackage{mathpazo}
\else
    \usepackage{fontspec}
\fi
\usepackage{graphicx}
\graphicspath{{./}}
\usepackage{tikz}
\usetikzlibrary{arrows.meta,calc}
\usepackage{amsmath}
\usepackage{bm}
\usepackage{amssymb}
\usepackage{caption,subcaption}
\usepackage[export]{adjustbox}
\usepackage{soul}
\usepackage{comment}
\usepackage{multirow}
\usepackage{xcolor}
\usepackage{authblk}
\usepackage{booktabs}
\usepackage{bbold}
\usepackage{relsize}
\usepackage[left=2cm, right=2cm, top=2cm]{geometry}
\usepackage{framed}
\usepackage[
backend=biber,
style=numeric,
sorting=none,
sortcites=true
]{biblatex}
\usepackage[hyperfootnotes=false]{hyperref}
\definecolor{urlcolor}{rgb}{0,.145,.698}
\definecolor{linkcolor}{rgb}{.71,0.21,0.01}
\definecolor{citecolor}{rgb}{.12,.54,.11}
\hypersetup{
breaklinks=true,
colorlinks=true,
urlcolor=urlcolor,
linkcolor=linkcolor,
citecolor=citecolor}
\usepackage{array}
\usepackage[title]{appendix}
\newcolumntype{P}[1]{>{\centering\arraybackslash}p{#1}}

\title{Plastic Flow Rule Based on Dislocation Link Velocity}

\author[1]{Hanfeng Zhai\textsuperscript{\dag}}
\author[1]{Sh. Akhondzadeh\textsuperscript{\dag}}
\affil[1]{Department of Mechanical Engineering, Stanford University, Stanford, CA, 94305, USA}
\author[2]{Ryan B. Sills}
\affil[2]{Department of Materials Science and Engineering, Rutgers University, Piscataway, NJ, 08854, USA}
\author[3]{Nicolas Bertin}
\affil[3]{Lawrence Livermore National Laboratory, Livermore, CA, 94550, USA}
\author[1]{Wei Cai\footnote{E-mail: \texttt{caiwei@stanford.edu}}}

\begin{document}
\maketitle
\begingroup
\renewcommand{\thefootnote}{\fnsymbol{footnote}}
\footnotetext[2]{These authors contributed equally to this work.}
\endgroup

\begin{abstract}
%By coarse-graining discrete dislocation dynamics (DDD) simulations, 
We develop a theory for the plastic shear rate $\dot{\gamma}_i$ on each slip system $i$ of face-centered cubic (FCC) single crystals, by integrating the shear rate contributions $\dot{\gamma}_i(l)$ from dislocation links of length $l$.
%
%The DDD simulations prescribe a linear dislocation mobility law, yet the resulting $\dot{\gamma}_i$ depends exponentially on the resolved shear stress $\tau_i$, a form usually associated with thermally activated motion; explaining this counterintuitive result is a central goal of this work.
%
A generalized Orowan equation expresses $\dot{\gamma}_i(l)$ in terms of the link density $n_i(l)$ and the average link velocity $v_i(l)$.
This model accurately reproduces the discrete dislocation dynamics (DDD) shear-rate data for all 12 slip systems across 99 loading orientations.
In particular, it explains why the plastic shear rate depends exponentially on the resolved shear stress $\tau_i$, as reported previously, in the absence of any thermally activated mechanism.
This dependence arises from the exponential tail of $n_i(l)$ filtered through the threshold-like velocity function $v_i(l)$.
Furthermore, $v_i(l)$ is well approximated by a single two-dimensional function $v(\tau_i,l)$ for all these slip systems.
The link velocity is therefore governed mainly by $\tau_i$ and $l$, with only a minor effect from the complex interactions between slip systems.
These interactions instead manifest themselves mainly through $n_i(l)$.
As a demonstration of the effect of the link-length distribution on the slip rate, we show that even when $\tau_i$ is limited to a narrow window, the shear rates still span two orders of magnitude.
The supply of long links is therefore the dominant factor controlling slip-system activity, outweighing the direct effect of $\tau_i$ on the link velocity.
\end{abstract}

%\clearpage

%\tableofcontents

%\clearpage

\section{Introduction}
\label{sec:intro}

Plastic deformation in crystalline materials is primarily driven by the motion of dislocations, which are line defects within the crystal lattice. Despite decades of progress in understanding the fundamental mechanisms of individual dislocations, a quantitative link remains missing between microscopic dislocation dynamics and macroscopic constitutive relations. A primary reason for this gap is that dislocations self-organize into highly complex network structures during plastic deformation. To bridge this divide, a dislocation-based theory of crystal plasticity must address two critical questions: (1) which quantifiable microstructural parameters of the dislocation network dictate plastic strength, and (2) how do these parameters evolve with plastic strain? While discrete dislocation dynamics (DDD) simulations provide the highly detailed data necessary to investigate these questions, systematic frameworks are still required to extract physical insights and derive constitutive laws from such massive datasets.
%
%Dislocations are line defects in crystals that interact through long-range stress fields and short-range junction reactions, so the link between microscopic dislocation motion and macroscopic stress--strain response during strain hardening is not available in a simple closed form~\cite{Kubin2013,bulatov2006computer,Madec2002_prl_junction_forest_harden,Madec2003_science}.
%
%Individual dislocation glide is the dominant mechanism for crystalline plastic flow under most loading conditions~\cite{taylor1934mechanism,orowan1934plasticity,polanyi1934lattice}.
%
%At the same time, obstruction, multiplication, and annihilation events make local driving stresses highly heterogeneous in space and time, so similar applied loads can produce different local velocities and path-dependent collective responses~\cite{devincre2008dislocation,papanikolaou2012quasi}.
%
%As a result, continuum stress--strain behavior should be viewed as a statistical aggregate of many microscopic events rather than as a direct one-to-one map from stress to a single microscopic state~\cite{roters2010overview,zaiser2019stochastic}.

\begin{figure}[htbp]
    \centering
    \includegraphics[width=0.40\linewidth]{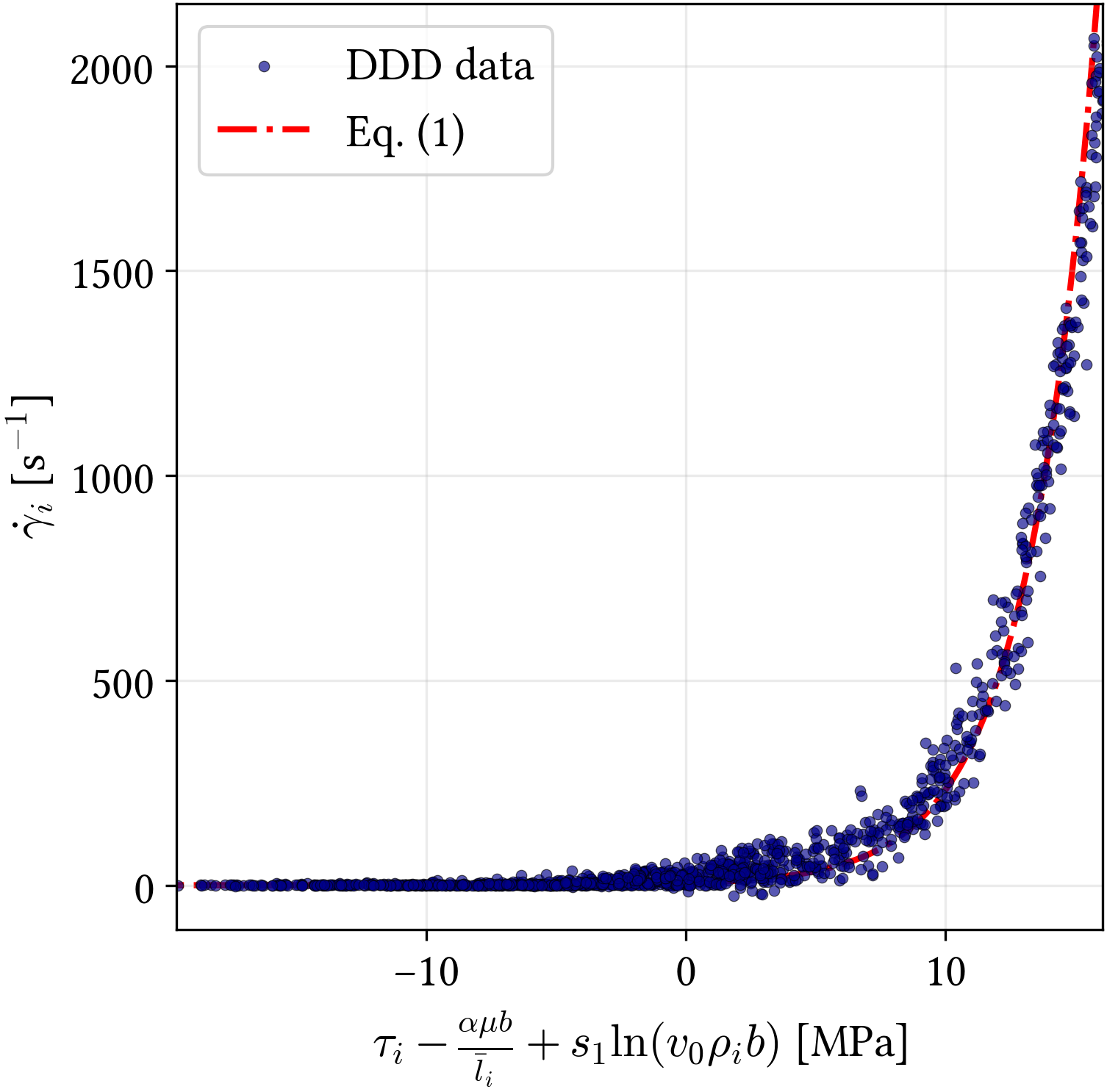}
    \caption{Comparison of flow-rule predictions against DDD data~\cite{JMPS1, akhondzadeh2021statistical}.
    Plastic shear rate on a given slip system as a function of the ``effective stress'' on that slip system.}
    \label{fig:gammaidot_pred_intro}
\end{figure}

To capture the well-established experimental observation that single-crystal responses heavily depend on loading orientation, crystal plasticity models need to be cognizant of individual slip systems~\cite{franciosi1980latent,peirce1983material}. 
For instance, face-centered cubic (FCC) crystals possess 12 primary slip systems, comprising 4 distinct slip planes with 3 slip directions each (Appendix~\ref{app:slip_systems}). Within these frameworks, the dislocation density on each slip system, $\rho_i$ (where $i=1,\cdots,12$), is commonly adopted as the primary microstructural parameter characterizing the dislocation network.
Such models require a constitutive flow rule that expresses the plastic shear rate, $\dot{\gamma}_i$, as a function of the resolved shear stress (from the applied load alone), $\tau_i$, on slip system $i$ and the dislocation densities across all slip systems.
%
%Throughout this paper, $\tau_i$ is from the applied load alone.
%
To move beyond purely phenomenological descriptions~\cite{hutchinson1976bounds}, we recently obtained a flow-rate -- stress relationship~\cite{JMPS1,MT2020} from a large database of DDD simulations.
%
%
%For crystal plasticity, this motivates a slip-system-level flow rule that predicts $\dot{\gamma}_i$ from resolved stress and internal variables that summarize the evolving microstructure~\cite{franciosi1980latent,peirce1983material}.
%
%In practice, this closure is evaluated repeatedly at each time step, so its structure strongly affects both predictive capability and numerical robustness.
%
%Earlier work on the same DDD database showed that an effective exponential relation can capture the stress--rate trend after coarse graining~\cite{JMPS1,MT2020}.
%
%The effective form is
We found that the DDD data can be reasonably well described by an exponential function (see Fig.~\ref{fig:gammaidot_pred_intro}),
\begin{equation}
    {\dot{\gamma}_i} = {\rho_i b}\, v_0\exp{\left[\frac{1}{s_1}\left(\tau_i-\left(\frac{\alpha \mu b}{\bar{l}_i}-\tau_0\right)\right)\right]},
    \label{eq:gamma_dot_exp}
\end{equation}
where $v_0$, $s_1$, $\tau_0$, and $\alpha$ are fitting coefficients, $\bar{l}_i$ is the average length of dislocation links on slip system $i$, $b$ is the magnitude of the Burgers vector and $\mu$ is the shear modulus.
The term $\alpha\mu b/\bar{l}_i$ represents the self stress of a dislocation link of length $\bar{l}_i$, which opposes the bowing out of the link between its end points.
Note that this constitutive relation utilizes two distinct parameters, $\rho_i$ and $\bar{l}_i$, per slip system to characterize the dislocation microstructure.

An exponential relationship between the resolved shear stress $\tau_i$ and the plastic shear rate $\dot{\gamma}_i$ is typical of thermally activated dislocation motion, where individual dislocation velocities depend exponentially on stress. 
This behavior is traditionally expected in crystals exhibiting high lattice resistance (Peierls barrier), such as semiconductors or body-centered cubic (BCC) metals~\cite{taylor1992thermally}. 
Conversely, glissile dislocations in face-centered cubic (FCC) metals experience negligible lattice resistance, and their velocities are governed instead by phonon drag~\cite{cai2004dislocation}. 
Consistent with this underlying physics, a linear mobility law was explicitly prescribed in the DDD simulations, entirely omitting thermally activated mechanisms. 
These considerations render the emergence of an exponential macroscopic flow-rate expression, Eq.~(\ref{eq:gamma_dot_exp}), from purely linear dislocation mobility inputs highly counterintuitive.
%
%However, when that relation is overlaid on slip-system-resolved data, substantial scatter remains, which indicates that stress-only closure is not sufficient when the underlying dislocation structure varies strongly among systems.

%\hl{(We want to explain this empirical form.  Why exponential form even if we don't have thermal activation in the model.  What is the physics behind it?)}

A major goal of this paper is to uncover the fundamental physics underlying Eq.~(\ref{eq:gamma_dot_exp}).  Specifically, we aim to explain why the flow rate $\dot{\gamma}_i$ is exponentially dependent on $\tau_i$, even though the motion of individual dislocation links is athermal (not thermally activated).  We hypothesize that the answer lies in the statistical properties of the dislocation links. 
We have previously shown that the link length distribution, $n_i(l)$, can be well described either by a single exponential function or the sum of two exponential functions~\cite{akhondzadeh2026link}.
Consequently, the exponential dependence of $\dot{\gamma}_i$ on $\tau_i$ may arise because a slip system $i$ requires sufficiently long links to achieve an appreciable strain rate.
%
%A major goal of this paper is to explain the origin of the exponential flow rule, Eq.~(\ref{eq:vEff_expTaueff}), in the cases where the motion of individual dislocations do not require thermal activation. 
%
For this purpose, we formulate a link-length based flow rule, using a generalized form of the Orowan equation ($\dot{\varepsilon} = \rho\,b\,v$).  Here, $\dot{\gamma}_i$ is defined as the integrated contribution of links across all lengths $l$, denoted as $\dot{\gamma}_i(l)$:

\begin{equation}    \dot{\gamma}_i=\int_0^{\infty}\dot{\gamma}_i(l)\,dl=\int_0^{\infty}n_i(l)\,{v_i}(l)\,b\,l\,dl
    \label{eq:Orowan_L}
\end{equation}
where $v_i(l)$ is the average velocity of a link of length $l$ on slip system $i$.
%subjected to the resolved shear stress $\tau_i$. 
%
This link velocity function $v_i(l)$ serves as the bridge between individual dislocation links and the macroscopic crystal plasticity.

In this paper, we show that on an active slip system $i$, the length-resolved plastic shear rate $\dot{\gamma}_i(l)$ peaks at a link length of a few times $\bar{l}_i$, mostly between $2\,\bar{l}_i$ and $5\,\bar{l}_i$, where $\bar{l}_i$ is the average link length on the slip system.
Given the exponential nature of the link length distribution $n_i(l)$, only about $14\%$ of the links are longer than $2\,\bar{l}_i$, and fewer than $2\%$ are longer than $4\,\bar{l}_i$.
The fact that such a small fraction of the total link population dominates the macroscopic flow rate confirms our hypothesis: the activation of a slip system relies entirely on the tail of the distribution.  Specifically, a rare population of long links that can easily bow out and multiply under the resolved shear stress $\tau_i$ is required for activating a slip system.
%
%This confirms the above hypothesis that the activation of a slip system $i$ requires a population of long links that can move under resolved shear stress $\tau_i$.

To quantify this behavior, the link velocity function $v_i(l)$ on slip system $i$ describes how the mobility of a link increases with its length.
We show that $v_i(l)$ on every slip system can be well described by a sigmoid function of $l$ controlled by three parameters $\theta_{i}^{(1)}$, $\theta_{i}^{(2)}$, and $\theta_{i}^{(3)}$.
The link velocity function $v_i(l)$, together with the link length distribution function $n_i(l)$, predicts the macroscopic slip rates (through Eq.~\eqref{eq:Orowan_L}) that agree well with the DDD dataset.
Furthermore, we find that $v_i(l)$ can be well approximated by a two-dimensional function $v(\tau_i, l)$, where $\tau_i$ is the resolved shear stress of slip system $i$.
This is a surprising result, because it implies that the dislocation link velocity on all 12 slip systems across 99 different loading orientations can be well approximated by a single two-dimensional function $v(\tau_i, l)$.
In other words, the complex interactions between slip systems have a relatively minor effect on the link velocities, compared to the effects of the resolved shear stress $\tau_i$ and link length $l$.
The link velocity function therefore offers a great simplification in describing the complex problem of slip-system activity during plastic deformation.

%Intuitively, the resolved shear stress $\tau_i$ is expected to be the primary factor dictating the profile of the link velocity function.  We may signify this effect by writing $v_i(l; \tau_i)$.
%
%In reality, our analysis reveals that $v_i(l)$ is also influenced by other factors, such as dislocation densities and resolved shear stresses on other slip systems.
%
%However, we find that the link velocity functions exhibit a remarkable degree of similarity across different slip systems and loading orientations. 
%
%Nonetheless, we found that the link velocity functions $v_i(l)$ do not vary too much among all slip systems and loading orientations.
%
%This is especially true if we limit the slip systems subjected to similar resolved shear stress $\tau_i$.
%
%This uniformity becomes particularly evident when evaluating the 73 distinct slip systems with resolved shear stress between 8 and 9~MPa.

%It is particularly instructive to consider the (73) slip systems whose $\tau_i$ is between $8$ and $9$~MPa.  Even though their $\tau_i$ values fall into a narrow range, their shear rates varies from $??$ to $??$, i.e. spans the entire range from being inactive to active.
%
The above framework also reveals a strong effect of the link-length distribution $n_i(l)$ on whether or not a slip system is activated.
As will be shown, this can be illustrated by restricting our attention to the $73$ slip systems, across various loading orientations, whose resolved shear stress $\tau_i$ lies between $8$ and $9$~MPa.
Even though their $\tau_i$ values fall within this tightly constrained range, their plastic shear rates vary by two orders of magnitude, from roughly 15 to 1500 s$^{-1}$.
They span the entire spectrum from inactive to highly active. 
Despite this vast divergence in macroscopic behavior, their underlying link velocity functions $v_i(l)$ remain very similar to each other. In fact, we may replace the individual velocity functions $v_i(l)$ for all 73 systems with their collective average $\bar{v}(l)$.
The resulting shear rates predicted by Eq.~(\ref{eq:Orowan_L}) still show good agreement with the DDD data, and faithfully capture the full transition from inactive to active slip behavior.

%The link velocity functions $v_i(l)$ for these slip systems are quite similar to each other.
%
%In fact, if we replace $v_i(l)$ for all these slip systems with their average $\bar{v}(l)$ in Eq.~(\ref{eq:Orowan_L}), the resulting prediction of the shear rate $\dot{\gamma}_i$ still agrees very well with the DDD data.
%
%Importantly, this mean-velocity prediction still faithfully captures the full transition from inactive to active slip behavior. 

The above results confirm that the empirically observed exponential flow rule, Eq.~(\ref{eq:gamma_dot_exp}), is rooted in the link-length-based ansatz, Eq.~(\ref{eq:Orowan_L}).
Specifically, the macroscopic exponential behavior emerges from a sensitive geometric interplay: the tail of the exponential link-length distribution $n_i(l)$ filtering through the threshold-like behavior of the sigmoidal velocity function $v_i(l)$. 
Because $v_i(l)$ can be well approximated by the single function $v(\tau_i, l)$, the complex interactions between slip systems manifest themselves more clearly in the link-length distribution $n_i(l)$ than in the link velocity.
It is through $n_i(l)$ that they exert the strongest influence on the slip rate $\dot{\gamma}_i$.

The remainder of this paper is organized as follows.
Section~\ref{sec:methods} describes the procedure to extract slip system and link information from DDD simulations, as well as the method for fitting the parameters for the link velocity functions.
%data extraction workflow, loading-orientation sampling, and slip-system-resolved observables.
%
Section~\ref{sec:results} presents the obtained link velocity functions and their ability to reproduce the shear rate on all slip systems compared to DDD data.
%representative orientation-dependent results, and Section~\ref{sec:velocity_function} introduces the velocity-function form, fitting strategy, and flow-rule comparison.
%
It also discusses the behavior of the link velocity function across all slip systems over all loading orientations.
Section~\ref{sec:discussion} discusses how the exponential flow rule, Eq.~(\ref{eq:gamma_dot_exp}), can be understood within the link-based framework.
It also highlights the importance of the tails in the link-length distribution for the activation of a slip system.
Section~\ref{sec:conclusion} summarizes our main findings and their implications for constitutive modeling of crystal plasticity.

\section{Methods}
\label{sec:methods}

\subsection{Data generation}

\begin{figure}[htbp]
    \centering
    \begin{subfigure}[t]{0.32\linewidth}
    \includegraphics[width=\linewidth]{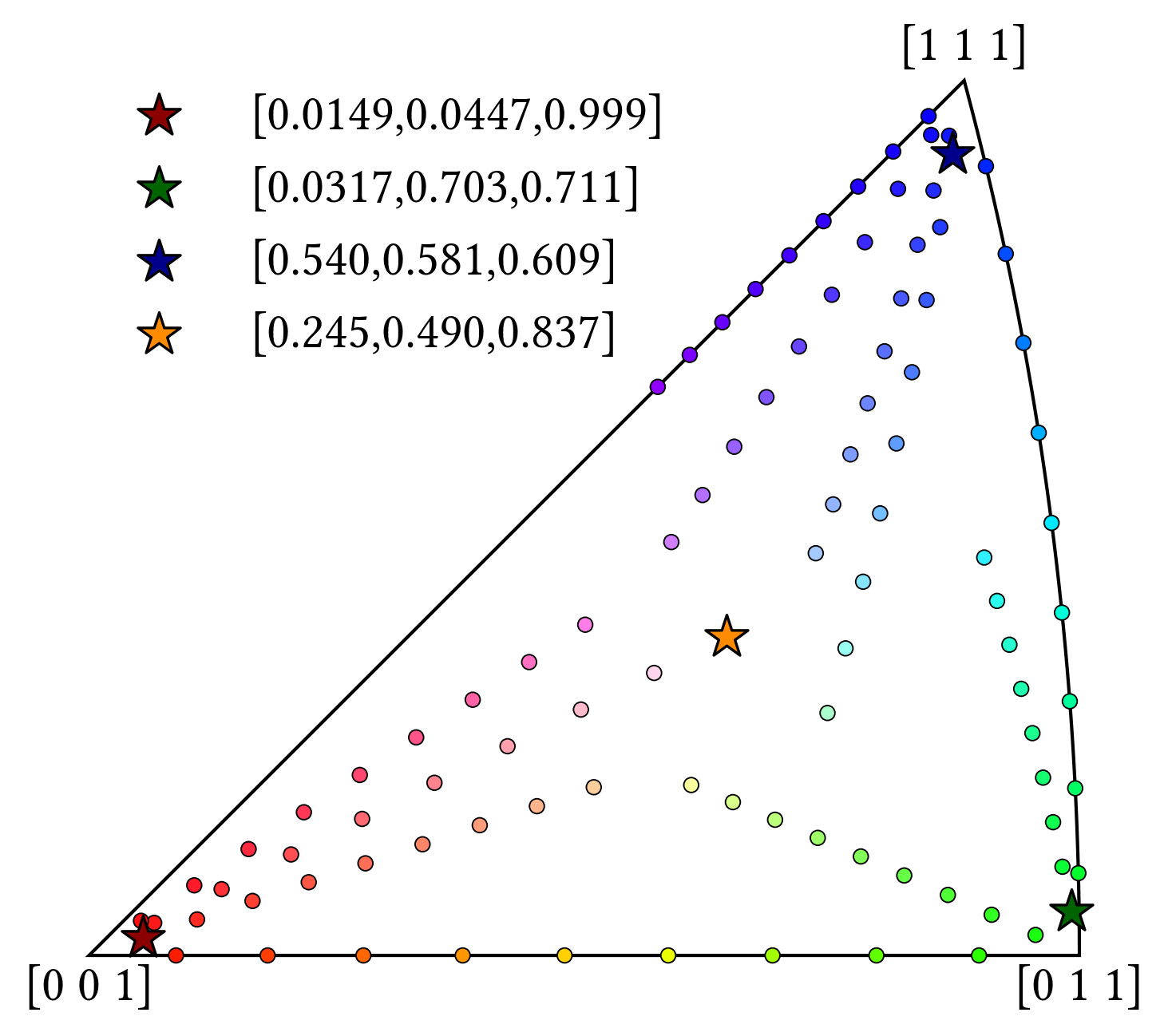}
    \caption{}
    \label{fig:stereo_triangle}
    \end{subfigure}\hspace{25pt}
    \begin{subfigure}[t]{0.40\linewidth}
    \includegraphics[width=\linewidth]{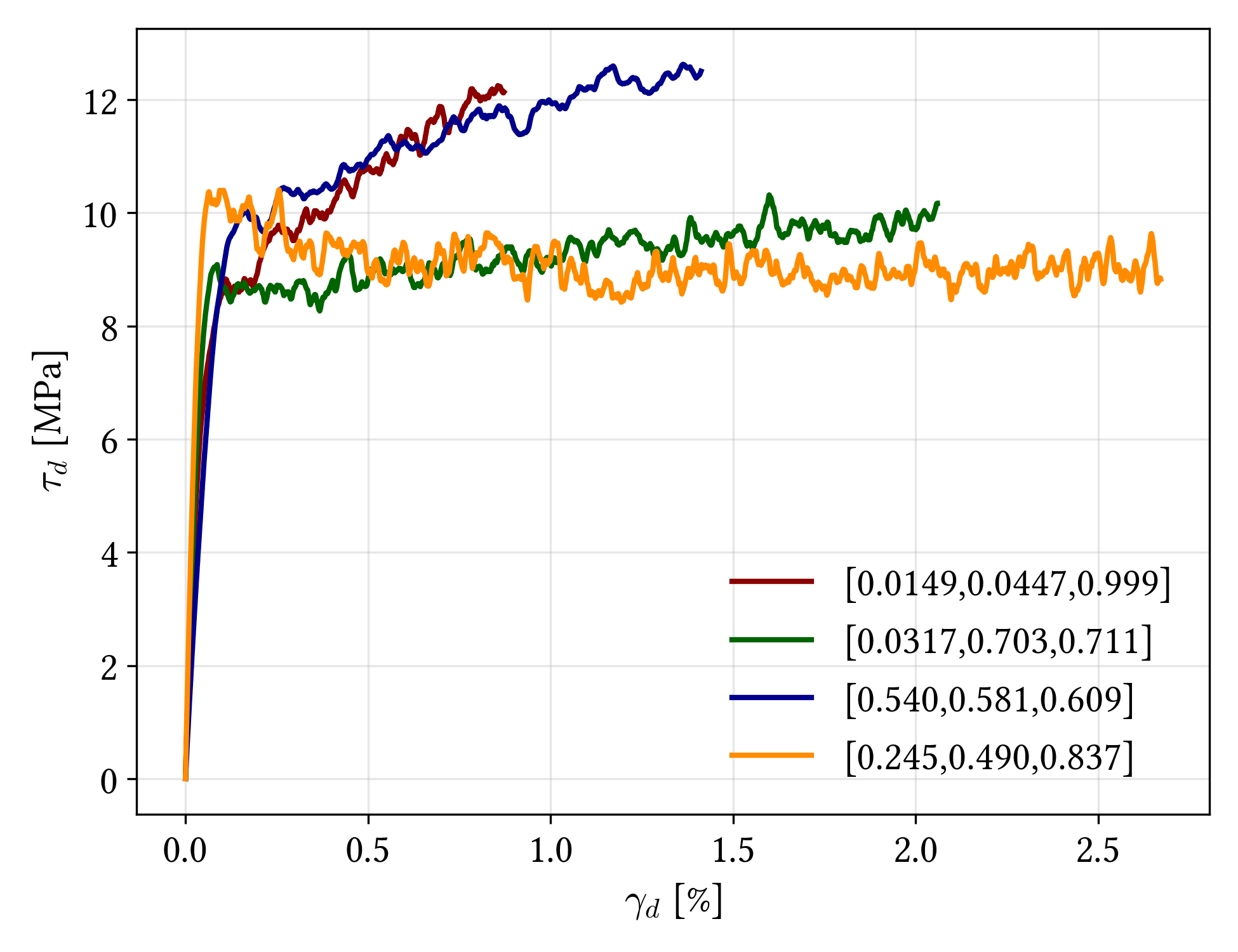}
    \caption{}
    \label{fig:stress_strain}
    \end{subfigure}
    \caption{(a) 99 loading orientations for DDD simulations on the stereographic triangle.
    Four selected orientations near $[001]$, $[011]$, $[111]$, and $[123]$ are marked as stars.
    (b) Shear stress--strain curves of the four selected loading orientations in (a).
    %$(\tau_d,\gamma_d)$ using the same colors.
    %
    %The figure highlights orientation-dependent hardening behavior that motivates orientation-aware constitutive closure.
    }
    \label{fig:stress_strain_stereo}
\end{figure}

We use data from discrete dislocation dynamics (DDD) simulations performed with the ParaDiS code for FCC copper~\cite{JMPS1,Arsenlis2007_msmse}, with properties: shear modulus $\mu=54.6$~GPa, Poisson ratio $\nu=0.324$, Burgers vector magnitude $b=0.255$~nm, and drag coefficient $B=15.6~\mu\mathrm{Pa}\cdot\mathrm{s}$.
Each simulation starts from the same randomly distributed dislocation lines
% \hl{(are all simulations starting from the same or different random configurations?)}
in a cubic cell of size $V=(15\,\mu\mathrm{m})^3$ and is subjected to uniaxial loading at a strain rate of $10^3~\mathrm{s}^{-1}$ along one of $99$ orientations sampled across the stereographic triangle (Figure~\ref{fig:stress_strain_stereo}(a)).
Different loading orientations give rise to different strain-hardening rates, as shown by the shear stress--strain curves in Figure~\ref{fig:stress_strain_stereo}(b).

The 12 slip systems of the FCC crystal are labeled in Schmid--Boas notation (Table~\ref{tab:slip_systems}).
Slip-system shear rates are extracted from the moving dislocation network by tracking the area swept by each link during a DDD time step~\cite{Arsenlis2007_msmse}.
%
%\hl{(Double check code if the following description is accurate.)}
%
Each link $k$, with length $l_k$, is assigned to the slip system $i$ defined by its Burgers vector and its slip plane.
Its swept area gives an incremental plastic strain, which is projected onto that slip system to obtain the slip increment $\Delta\gamma_i^{(k)}$ of that link.
Dividing by the DDD time step $\Delta t$ gives the contribution $\dot{\gamma}_i^{(k)}=\Delta\gamma_i^{(k)}/\Delta t$ of that link to the shear rate.
These contributions are binned by normalized link length $l/\bar{l}_i$.
The quantity accumulated in each bin is $\dot{\gamma}_i(l)\,\Delta l$, where $\Delta l$ is the physical bin width, and
$\dot{\gamma}_i(l)$ is the integrand in Eq.~\eqref{eq:Orowan_L}.
%
%\hl{Summing the binned values over all bins recovers the total shear rate $\dot{\gamma}_i$.}
%
The detailed construction is given in Appendix~\ref{app:plastic_strain}.
%
% \hl{(The description of how $\dot{\gamma}_i$ and $\dot{\gamma}_i(l)$ are calculated still sounds strange.)}

\begin{comment}
The different trends in the stress--strain curves shown in Figure~\ref{fig:stress_strain_stereo}(b) originate from how the dislocation-network topology interacts with different loading orientations.
%
Slip-system shear rates are extracted from the moving dislocation network by tracking the area swept by each link during a DDD time step~\cite{Arsenlis2007_msmse}.
%
The swept area of each segment defines an incremental plastic strain tensor, which is projected onto each slip system $i$ and summed over its links to give the shear rate $\dot{\gamma}_i$.
%
Binning the link-level contributions by normalized link length $l/\bar{l}_i$ then yields the length-resolved shear rate $\dot{\gamma}_i(l)$ that enters Eq.~\eqref{eq:Orowan_L}.
%
The detailed construction of $\dot{\gamma}_i$ and $\dot{\gamma}_i(l)$ from the swept areas is given in Appendix~\ref{app:plastic_strain}.
\end{comment}

\begin{figure}[ht]
    \centering
    \begin{subfigure}[t]{0.4\linewidth}
    \includegraphics[width=\linewidth]{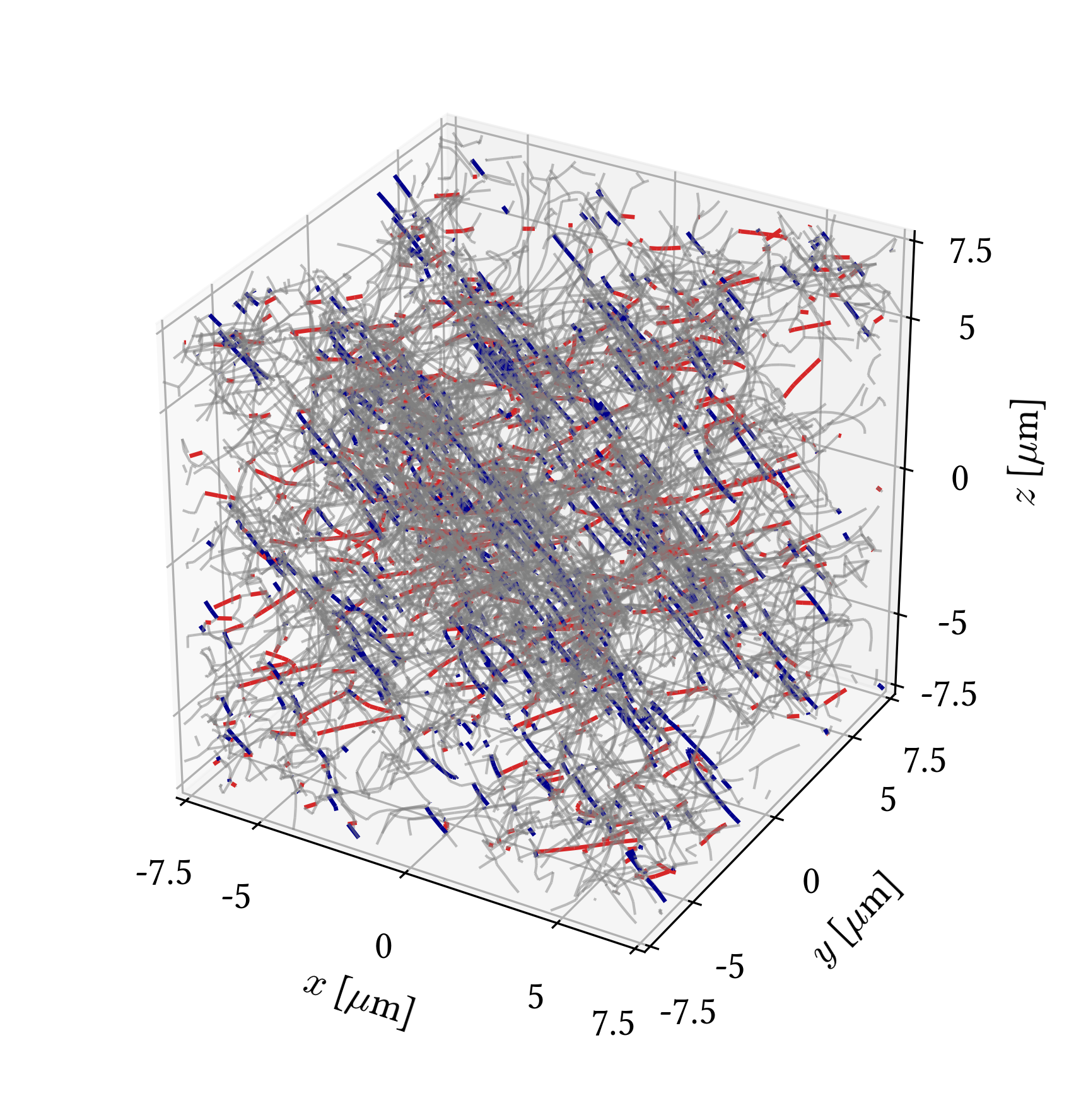}
    \caption{}
    \label{fig:network_slip}
    \end{subfigure}\hspace{20pt}
    \begin{subfigure}[t]{0.4\linewidth}
    \includegraphics[width=\linewidth]{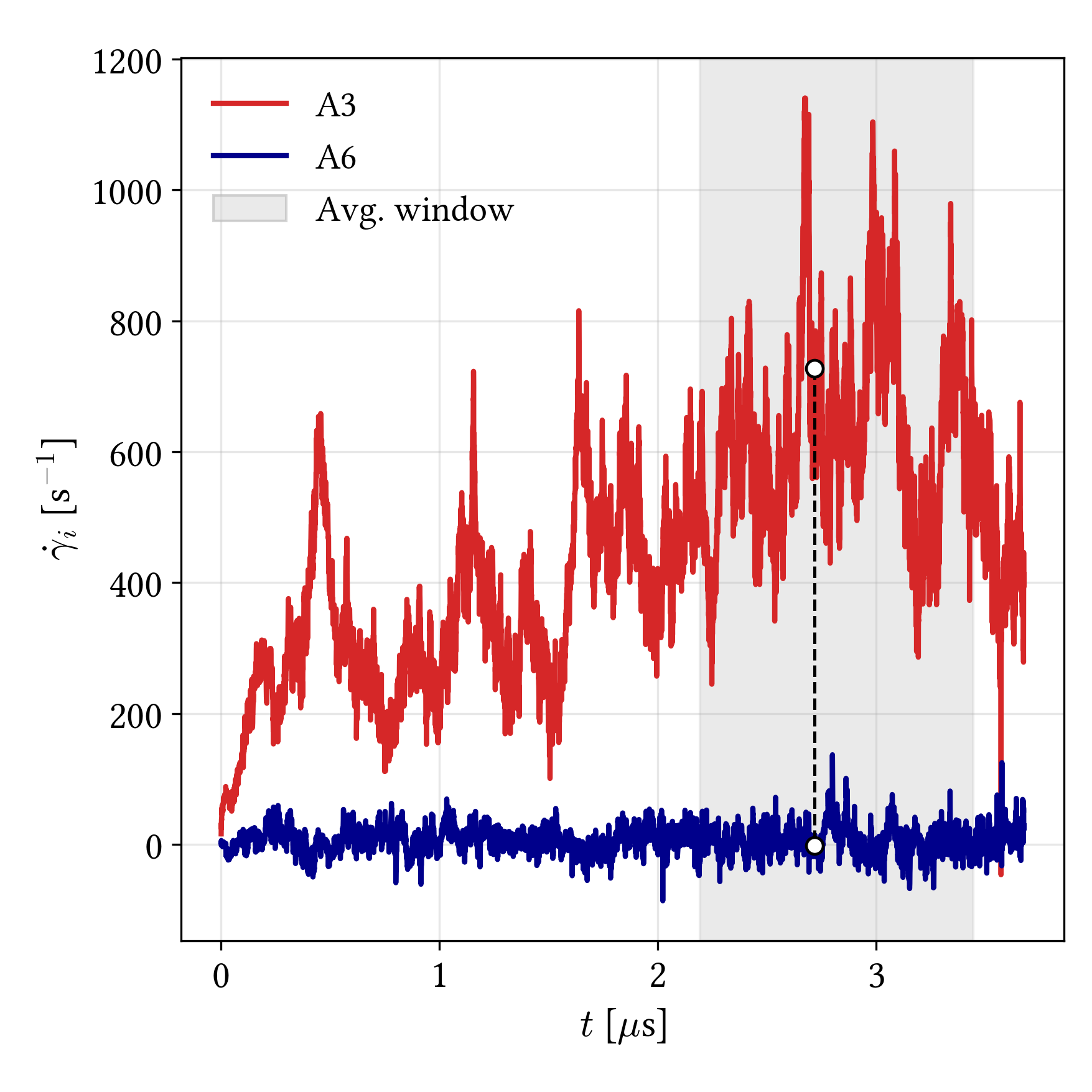}
    \caption{}
    \label{fig:gammaidot}
    \end{subfigure}
    \caption{Slip-system-level data extraction example for loading orientation $[0.0149,0.0447,0.999]$ (close to $[001]$).
    % Same orientation as Figure~\ref{fig:close_001}, confirmed by the co-author.
    % Full-precision loading direction (from main_zhai.tex):
    %   [0.014899685704935852, 0.04469905711480755, 0.9988893800911797]
    %
    %\hl{keep the dots and draw the dash line in between the 2 dots -- dashline indicate where (a) is taken.}
    %
    % \hl{(Is this the same orientation as in Figure.~4?) Yes}
    %
    (a) The dislocation network with slip systems $i=2$ and $3$ (A3 and A6) highlighted and all other systems grayed out.
    % \hl{(Change x, y, z units to $\mu$m instead of $b$.)}
    %
    (b) The corresponding $\dot{\gamma}_i(t)$ for the same two systems.
    The marked dots are at $t = 2.72\,\mu\mathrm{s}$, the network snapshot time in (a).
    %
    % \hl{(Remove the dashed line -- it attracts too much attention.  Just put a dot on the $t$ axis and on the caption say that the dot represents the $t = 2.72 \mu$s corresponding to (a).)}
    % The contrast between highlighted and grayed-out systems motivates slip-resolved constitutive modeling. 
    % \hl{(This kind of sentence does not belong to figure caption.  Figure caption only describes what the curves represent.)}
    %
    % \hl{Add shaded region to indicate from where to where is the shear rate averaged.}
    }
    \label{fig:network_gammaidot}
\end{figure}

% From DDD we obtain $\dot{\gamma}_i(l)$ by binning swept-area shear-rate contributions of moving links by link length, and $n_i(l)=N_i(l)/(V\,\Delta l)$ from link histograms, with both averaged over a strain window near the end of each run (the snapshot time marked in Fig.~\ref{fig:network_gammaidot}).
%
% Integrating $\dot{\gamma}_i(l)$ over $l$ recovers $\dot{\gamma}_i$ as in Eq.~\eqref{eq:Orowan_L}, and inverting the integrand gives $v_i(l)=\dot{\gamma}_i(l)/[n_i(l)\,b\,l]$ (with $v_i(l)\equiv\bar{v}_i(\tau_i,l)$).

% \hl{(This paragraph repeats previous paragraph.)}
Figure~\ref{fig:network_gammaidot} illustrates the slip-resolved shear-rate extraction for two slip systems of a near-$[001]$ loading orientation.
Figure~\ref{fig:network_gammaidot}(a) shows one network snapshot with the links on slip systems $i=2$ and $3$ (A3 and A6) highlighted (Table~\ref{tab:slip_systems}).
Figure~\ref{fig:network_gammaidot}(b) shows the corresponding $\dot{\gamma}_i(t)$ evolution for the same two systems.
The lower shear rate of system A6 relative to A3 is correlated with its lower Schmid factor.
The shear rate $\dot{\gamma}_i$ and its length-resolved counterpart $\dot{\gamma}_i(l)$ are extracted on each slip system $i$, and averaged over a strain window selected to improve statistics (shaded region in Figure~\ref{fig:network_gammaidot}(b)).
The link-length distribution $n_i(l)$ is averaged over the same window.

\subsection{Link velocity function}
\label{sec:velocity_fit}

%The slip-system shear rates in Figure~\ref{fig:stress_strain_stereo}(c) show that A3 produces a much larger contribution than A6 in the strain-hardening regime, indicating that the shear-rate response depends strongly on the link-sweep velocity induced by the resolved shear stress.
%
%Motivated by this observation, we therefore construct a velocity function from the DDD simulations to describe how the average link velocity varies with the physical link length $l$.
%
An estimate of the link velocity function $v_i(l)$ follows directly from the DDD data by inverting the integrand of Eq.~\eqref{eq:Orowan_L}, that is, by dividing $\dot{\gamma}_i(l)$ by $n_i(l)\,b\,l$.
The resulting profiles show that short links move relatively slowly, whereas longer links move faster and appear to approach an asymptotic velocity.
Motivated by this trend, we adopt a sigmoidal form for $v_i(l)$ and determine its parameters by fitting to the DDD shear-rate data $\dot{\gamma}_i(l)$,
%The link velocity data from DDD reveal a sigmoidal transition: short links remain relatively slow, whereas longer links approach a higher asymptotic velocity.
%
%Motivated by this trend, we represent a universal velocity function shared by all slip systems,
%\begin{equation}
%\begin{aligned}
%    \bar{v}(l; \tau)=\theta^{(3)}(\tau)\left[\sigma\!\left(\frac{l}{\theta^{(2)}(\tau)}-\theta^{(1)}(\tau)\right)-\sigma\!\left(-\theta^{(1)}(\tau)\right)\right],\quad
%    \text{where}\,\,\sigma(x)=\frac{1}{1+e^{-x}}.
%    \label{eq:ViL_fit}
%\end{aligned}
%\end{equation}
%
\begin{equation}
\begin{aligned}
    v_i(l)=\theta_i^{(3)}\tau_i \left[\sigma\!\left(\frac{l}{\theta_i^{(2)}}-\theta_i^{(1)}\right)-\sigma\!\left(-\theta_i^{(1)}\right)\right],\quad
    \text{where}\,\,\sigma(x)=\frac{1}{1+e^{-x}}.
    \label{eq:ViL_fit}
\end{aligned}
\end{equation}
Here $-\theta_i^{(1)}$ selects the point on the $\sigma(x)$ curve that serves as the origin, $l=0$, of the velocity function.
The parameter $\theta_i^{(2)}$ sets the length scale over which $v_i(l)$ rises toward its asymptotic value, and $\theta_i^{(3)}\tau_i$ the overall velocity scale.
The subtraction of $\sigma(-\theta_i^{(1)})$ is introduced because we assume that a link of vanishing length is immobile, $v_i(0)=0$.
By Eq.~\eqref{eq:ViL_fit}, we also assume that $v_i(l)=0$ for all $l$ when $\tau_i=0$, so that links do not move in the absence of a resolved shear stress.
The adequacy of this functional form is assessed a posteriori, through the quality of the resulting fits to the DDD data.

%Each of $\theta^{(1)}(\tau)$ and $\theta^{(2)}(\tau)$ is a logistic in resolved shear stress,
%\begin{equation}
%    \theta^{(k)}(\tau)=c_k+a_k\,\sigma\!\left(\frac{\tau-\tau_{0,k}}{w_k}\right),\qquad k=1,2,
%    \label{eq:theta_tau_logistic}
%\end{equation}
%
%while the amplitude is affine,
%\begin{equation}
%    \theta^{(3)}(\tau)=c_3+b_3\,\tau.
%    \label{eq:theta3_affine}
%\end{equation}
%
%Because a dislocation must be immobile at vanishing load, $\bar{v}(l;0)=0$ for every $l$, which holds if and only if $\theta^{(3)}(0)=0$.
%
%Each individual amplitude is written $\theta^{(3)}_i=u_i\tau_i$ with $u_i\ge 0$, so that rest condition holds on every slip-system curve.
%
%The shared function is shifted $\theta^{(3)}(\tau)\leftarrow\theta^{(3)}(\tau)-\theta^{(3)}(0)$, which reduces Eq.~\eqref{eq:theta3_affine} to $\theta^{(3)}(\tau)=b_3\tau$.

The parameters of all slip-system curves are fitted together in a single optimization.
Each curve $i$ (one loading orientation combined with one slip system) carries its own triplet $\bm{\theta}_i=(\theta_i^{(1)},\theta_i^{(2)},\theta_i^{(3)})$.
%
%Because a dislocation must be immobile at vanishing load, each individual amplitude is written
%\begin{equation}
%    \theta_i^{(3)}=u_i\,\tau_i,
%    \qquad u_i\ge 0,
%    \label{eq:theta3_ray}
%\end{equation}
%so that $v_i(l)=0$ for all $l$ when $\tau_i=0$.
%
The complete set $\bm{\theta}=\{\bm{\theta}_i\}$ is obtained by minimizing
\begin{equation}
    \mathcal{L}(\bm{\theta})
    = \mathcal{L}_{\mathrm{data}}(\bm{\theta})
    + \lambda_{\mathrm{reg1}}\,\mathcal{L}_{\mathrm{reg1}}(\bm{\theta})
    + \lambda_{\mathrm{reg2}}\,\mathcal{L}_{\mathrm{reg2}}(\bm{\theta}),
    \label{eq:L_velocity_fit}
\end{equation}
with $\lambda_{\mathrm{reg1}}=10^{-2}$ and $\lambda_{\mathrm{reg2}}=1$.
The data term $\mathcal{L}_{\mathrm{data}}$ is the 
%\hl{curve-weighted}
mean-squared error between the predicted and the DDD length-resolved shear rates.
\begin{equation}
    \mathcal{L}_{\mathrm{data}}(\bm{\theta})
    =\frac{1}{N}\sum_{i}
    \left\langle
    \Big[\big(\dot{\gamma}_i^{\rm DDD}(l)-\dot{\gamma}_i^{\rm pred}(l)\big)\,\Delta l\Big]^{2}
    \right\rangle_{l\in\mathcal{G}_l},
    \label{eq:L_data}
\end{equation}
where $N$ is the number of curves, $\mathcal{G}_l$ is a fixed grid of link lengths, and $\langle\cdot\rangle_{l\in\mathcal{G}_l}$ is the mean over that grid.
For curve $i$, the prediction is the integrand of Eq.~\eqref{eq:Orowan_L} evaluated bin by bin,
\begin{equation}
    \dot{\gamma}_i^{\mathrm{pred}}(l) = n_i(l)\,v_i(l)\,b\,l\,
    \label{eq:dgamma_pred}
\end{equation}
where $n_i(l)$ is the link-length distribution of curve $i$, prefitted to the double-exponential form of Ref.~\cite{akhondzadeh2026link}, and $\Delta l$ is the bin width.
The fit is therefore performed on the shear-rate data rather than on velocity samples, which are considerably noisier.

The regularization term $\mathcal{L}_{\mathrm{reg1}}$ couples curves at nearby resolved shear stress, so that velocity profiles do not vary erratically from one curve to the next,
\begin{equation}
    \mathcal{L}_{\mathrm{reg1}}(\bm{\theta})
    = \frac{\sum_{l\in\mathcal{G}_l}\sum_{i<j} w_{ij}\,
            \big[v_i(l)-v_j(l)\big]^{2}}
           {\sum_{i<j} w_{ij}+\epsilon},
    \qquad
    w_{ij}=K_{\sigma,R}(|\tau_i-\tau_j|),
    \label{eq:L_reg}
\end{equation}
where $K_{\sigma,R}$ is a Gaussian kernel of width $\sigma$ truncated at radius $R$, and $\varepsilon$ guards the denominator.
%
%\hl{Proposed form for the (third) term in the loss function}
%
Similarly, 
\begin{equation}
    \mathcal{L}_{\mathrm{reg2}}(\bm{\theta})
    = \frac{\sum_{k\in\{1,2,3\}}\sum_{i<j} w_{ij}\,
            \big[\theta^{(k)}_i-\theta^{(k)}_j\big]^{2}}
           {\sum_{i<j} w_{ij}+\epsilon},
    \label{eq:L_reg2}
\end{equation} 
is introduced to prevent the fitted parameters from varying erratically from one curve to the next.
%
%Because the amplitude is proportional to $\tau_i$, the $k=3$ term compares the stress-normalized velocity scales $u_i$ rather than the raw amplitudes $\theta_i^{(3)}$.
%
Both regularization terms reduce overfitting to the noise in the DDD data.
They are justified a posteriori by the quality of the resulting fit, which confirms that they are weak enough not to distort the fitted velocity functions.

%{\color{red}no longer needed}
%\hl{To enforce}
%the physical requirement that links do not move in the absence of resolved shear stress, one synthetic curve per loading orientation is added at $\tau_i=0$ with $\theta_i^{(3)}=0$, so that $v_i(l)=0$ for all $l$.
%
%These curves are held fixed during the optimization.
%
%They act on the real curves through $\mathcal{L}_{\mathrm{reg1}}$ and $\mathcal{L}_{\mathrm{reg2}}$.
%
%\hl{(No longer applicable -- the new Eq.(3) already enforces this condition!)}

Despite the explicit factor $\tau_i$ in Eq.~\eqref{eq:ViL_fit}, the parameters $\bm{\theta}_i$ themselves carry no prescribed dependence on the resolved shear stress.
The coupling between curves at similar $\tau_i$ through the regularization terms is therefore what transmits statistical information among them.
We minimize Eq.~\eqref{eq:L_velocity_fit} with Adam and automatic differentiation~\cite{adam_optim,torch}, using a learning rate of $10^{-2}$ for up to $5000$ iterations.
We retain the best-loss iterate, and stop early if the loss does not improve for $300$ consecutive iterations.
%
%The full loss, the kernel, and all hyperparameters are given in Appendix~\ref{app:gaussian-tau-coupling}.

%
% --- retained from the Overleaf version (degree-4 polynomial theta(tau)) ---
%Each shared parameter is a degree-4 polynomial in resolved shear stress,
%\begin{equation}
%    \theta^{(k)}(\tau)=\sum_{p=0}^{4}\varphi_{kp}\left(\frac{\tau}{12}\right)^{p},\qquad k=1,2,3.
%    \label{eq:theta_tau_poly}
%\end{equation}

\section{Results}
%\section{Results and discussions}
\label{sec:results}

%\subsection{Distributions on shear rate}
\subsection{Link length-dependent contribution to shear rate}

\begin{figure}[htbp]
    \centering
    \includegraphics[width=\linewidth]{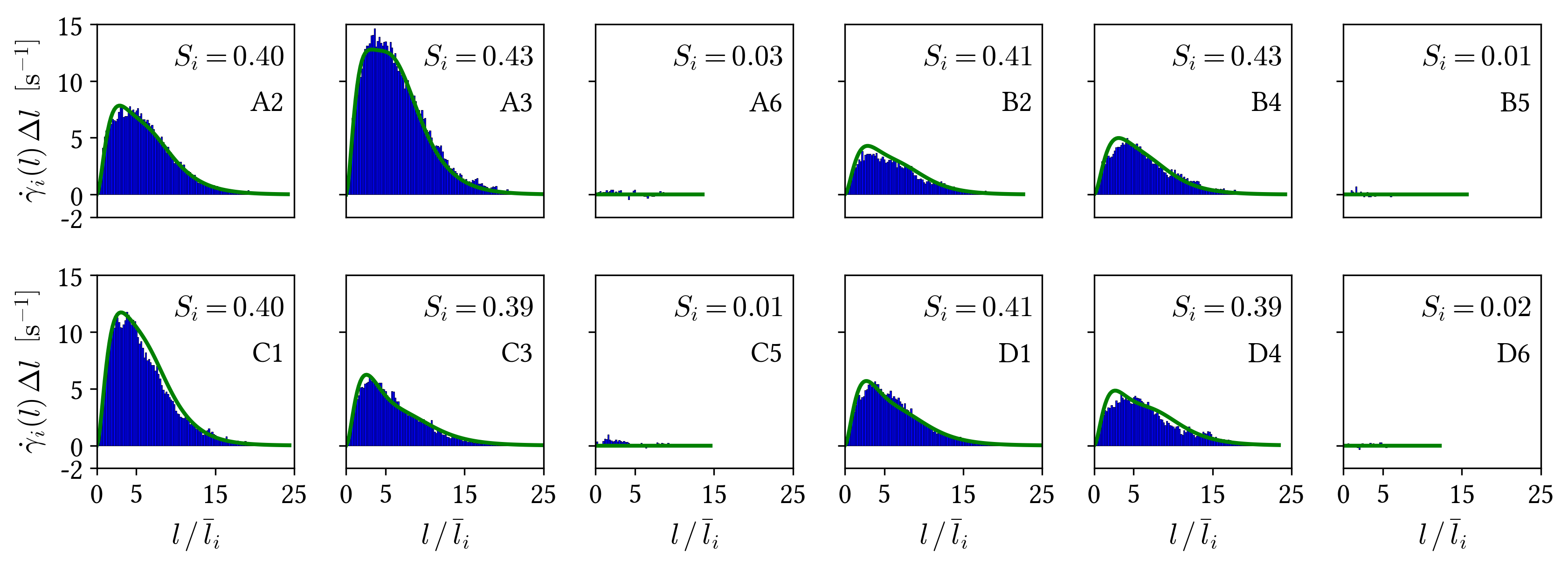}
    \caption{Length-resolved shear-rate distributions $\dot{\gamma}_i(l)\Delta l$ for loading orientation $[0.0149,0.0447,0.999]$ (close to $[001]$).
    % Full-precision loading direction (from main_zhai.tex):
    %   [0.014899685704935852, 0.04469905711480755, 0.9988893800911797]
    %
    Each panel corresponds to one slip system.
    Blue histograms are DDD-extracted data and green curves are predictions from the fitted velocity-function framework.
    %
    %\hl{(TODO: the blue histogram stores the per-bin sum $\dot{\gamma}_i(l)\,\Delta l$, whereas the green curve is the density $n_i(l)\,v_i(l)\,b\,l$.  Ask Hanfeng which of the two the plotting script rescales, with $\Delta l=0.2$.  Then fix the $y$-axis label here and in Figures 5--7.)}
    %
    %The comparison shows accurate recovery of slip-system-resolved distribution shape and amplitude.
    }
    \label{fig:close_001}
\end{figure}

\begin{figure}[htbp]
    \centering
    \includegraphics[width=\linewidth]{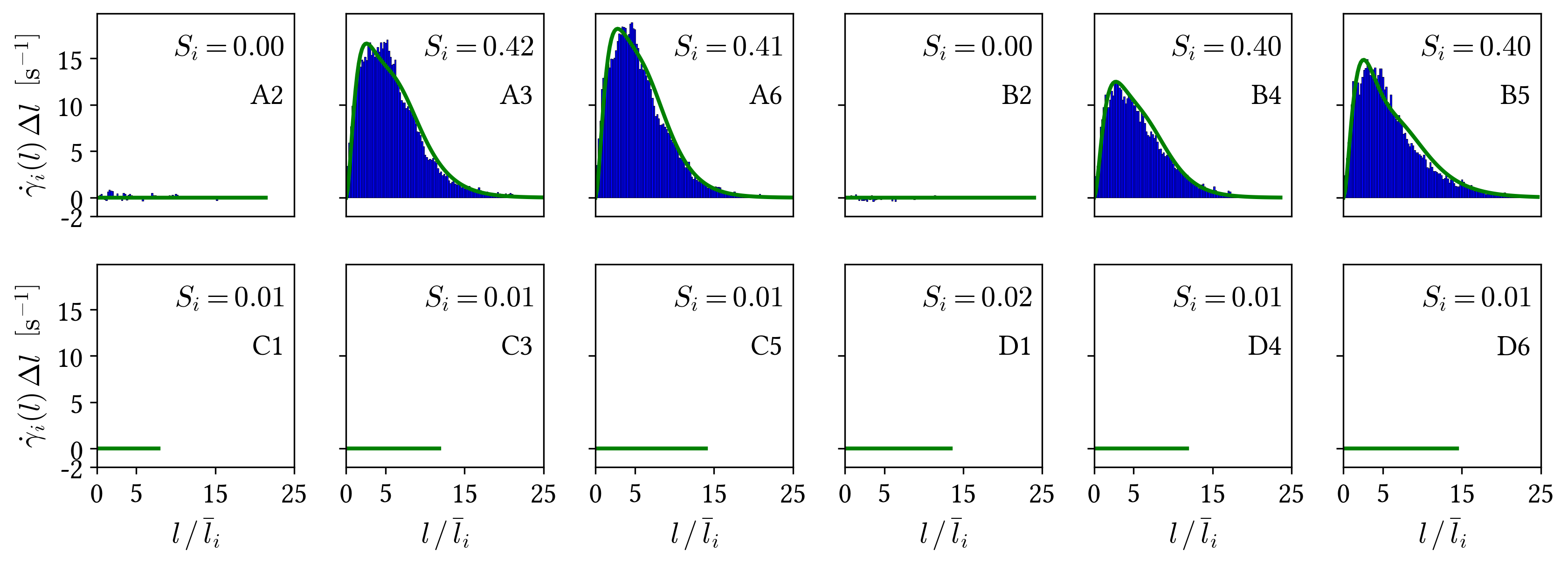}
    \caption{Length-resolved shear-rate distributions $\dot{\gamma}_i(l)\Delta l$ for loading orientation $[0.0317,0.703,0.711]$ (close to $[011]$).
    % Full-precision loading direction (from main_zhai.tex):
    %   [0.03167479881294172, 0.7027815341425322, 0.7107002338457675]
    %
    Symbols have the same meaning as in Figure~\ref{fig:close_001}.
    %
    %Blue histograms show DDD data and green curves show fitted predictions.
    %
    %The model captures orientation-dependent redistribution of slip activity while preserving quantitative agreement across systems.
    }
    \label{fig:close_011}
\end{figure}

\begin{figure}[htbp]
    \centering
    \includegraphics[width=\linewidth]{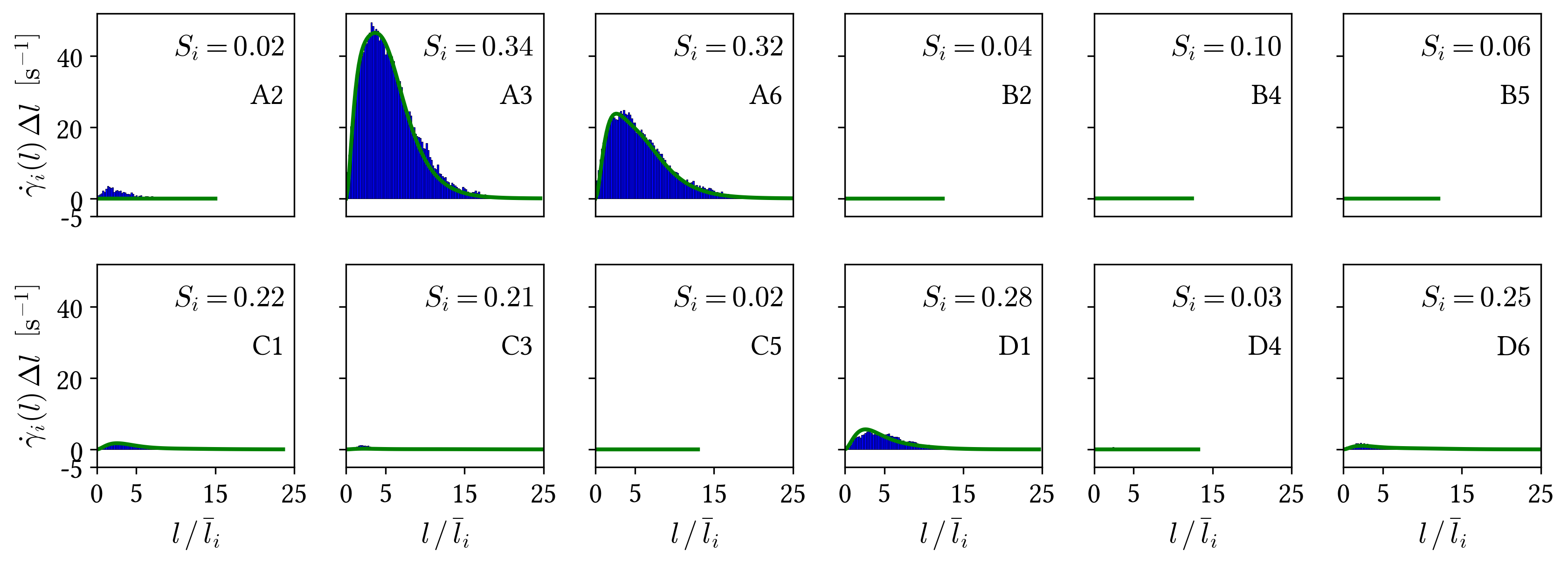}
    \caption{Length-resolved shear-rate distributions $\dot{\gamma}_i(l)\Delta l$ for loading orientation $[0.500,0.584,0.640]$ (close to $[111]$).
    % Full-precision loading direction (from main_zhai.tex):
    %   [0.4996632999822456, 0.5838162042065248, 0.6399181403560442]
    %
    Symbols have the same meaning as in Figure~\ref{fig:close_001}.
    %
    %Blue histograms denote DDD-extracted values and green curves denote model fits.
    %
    %Agreement across slip systems indicates that the fitted velocity form captures orientation-sensitive redistribution over link length.
    }
    \label{fig:close_111}
\end{figure}

\begin{figure}[htbp]
    \centering
    \includegraphics[width=\linewidth]{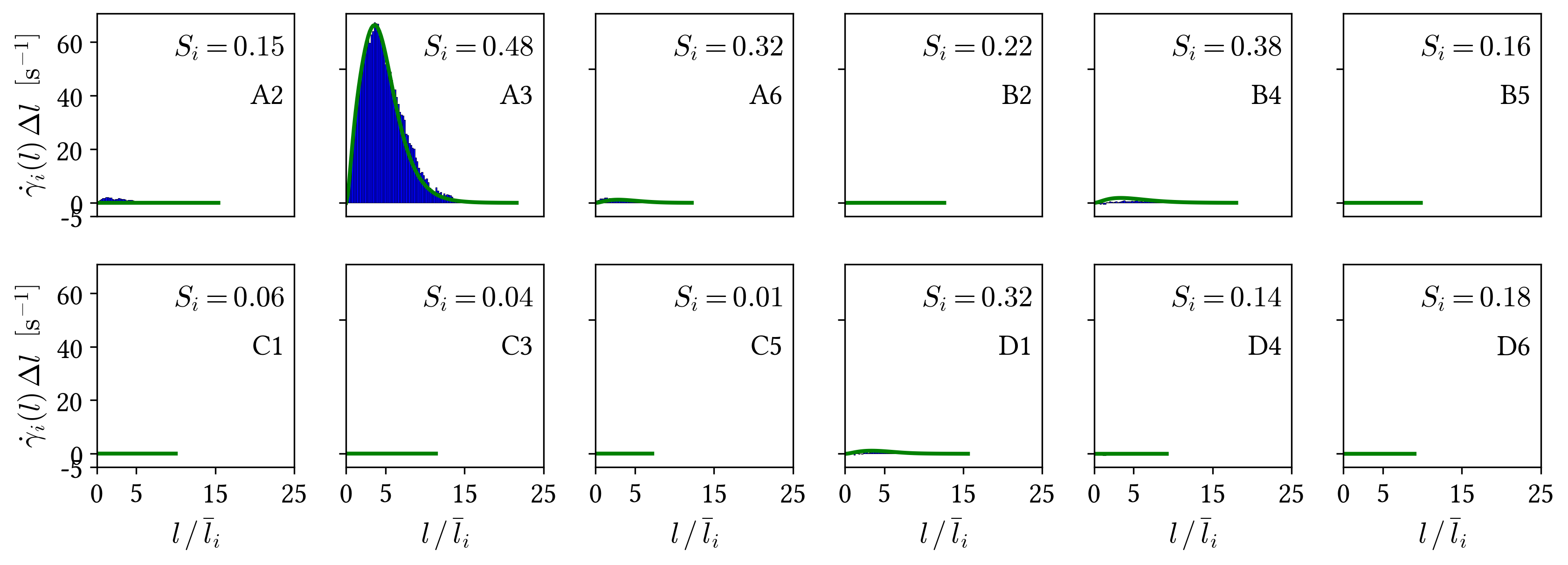}
    \caption{Length-resolved shear-rate distributions $\dot{\gamma}_i(l)\Delta l$ for loading orientation $[0.245,0.490,0.837]$ (close to $[123]$).
    % Full-precision loading direction (from main_zhai.tex):
    %   [0.24494448377169142, 0.48988896754338285, 0.8366665998766164]
    %
    Symbols have the same meaning as in Figure~\ref{fig:close_001}.
    %
    %Blue histograms represent DDD data and green curves represent fitted predictions.
    %
    %This case provides a central-triangle reference and confirms the same intermediate-length-dominant trend seen in other orientation neighborhoods.
    }
    \label{fig:close_123}
\end{figure}

Figure~\ref{fig:close_001} shows $\dot{\gamma}_i(l)$ distributions for a loading orientation close to $[001]$.
The fitted curves (green) are obtained by combining the measured link-length distribution $n_i(l)$ with the postulated stress-and-length velocity function $v_i(l)$, multiplied by $b\,l$.
They reproduce both the peak position and the relative activity level across slip systems.
Figure~\ref{fig:close_001} demonstrates that even though short links are the most numerous, the dominant shear-rate contribution comes from intermediate lengths where the product of link abundance and link mobility is largest.
For the eight active slip systems in Figure~\ref{fig:close_001}, the peak of the fitted $\dot{\gamma}_i(l)$ lies between $l/\bar{l}_i=2.4$ and $3.2$.
%
%\hl{(Not consistent with your plots for all loading orientations -- there is a range of values.)}
%
For this loading orientation, the fitted curves agree closely with the DDD data on every slip system in Figure~\ref{fig:close_001}.
They accurately reproduce the widely different magnitudes of the slip rate across the 12 slip systems.
This confirms a posteriori the length-resolved ansatz of Eq.~\eqref{eq:Orowan_L}, together with the sigmoidal form postulated in Eq.~\eqref{eq:ViL_fit} and the fitting procedure based on Eq.~\eqref{eq:L_velocity_fit}.
Two assumptions are built into the fit: that a link of vanishing length is immobile, $v_i(0)=0$, and that slip systems at similar resolved shear stress share similar velocity functions.
Neither introduces appreciable error relative to the DDD data.
The same quality of agreement is obtained for the other loading orientations, as shown in Figures~\ref{fig:close_011}--\ref{fig:close_123}.
%
%This directly validates the length-resolved mechanism encoded in Eq.~\eqref{eq:Orowan_L}.

Figure~\ref{fig:close_001} shows that the slip activity on a slip system is correlated with the Schmid factor.  As expected, the eight slip systems with Schmid factor near or above $0.4$ are all highly active, whereas the remaining four, with Schmid factor at or below $0.03$, carry negligible shear rate.
This correlation holds only within a given loading orientation.  We shall see later that, across different loading orientations, the Schmid factor alone does not determine whether a slip system is highly active.
Even among these eight active systems, the magnitude of the shear rate is not strictly ordered by the Schmid factor.
For example, C1 ($S=0.40$) is more active than B4 ($S=0.43$), despite B4's larger Schmid factor.

Figure~\ref{fig:close_011} shows the $\dot{\gamma}_i(l)$ distributions for a loading orientation near $[011]$.
%
%Figure~\ref{fig:close_111} examines a case close to $[111]$, where slip activity is redistributed again.
%
For this orientation, the four slip systems (A3, A6, B4, B5) with Schmid factor near or above $0.4$ are highly active, whereas the other eight, with Schmid factor at or below $0.02$, carry negligible shear rate.
Similar to Figure~\ref{fig:close_001}, the peak contribution to the shear rate still comes from links of intermediate length.
For the four active systems the peak lies between $l/\bar{l}_i=2.4$ and $2.7$.
The prediction $\dot{\gamma}_i(l)=n_i(l)\,v_i(l)\,b\,l$ still agrees very well with the DDD data.
%
%Figure~\ref{fig:close_011} shows that accuracy is not tied to one orientation family and remains consistent when slip activation changes.
%
%This behavior is achieved because orientation dependence enters through measured $n_i(l)$ and $\tau_i$ while the same calibrated functional form is retained.
%
%In particular, low-Schmid-factor systems remain at very low shear-rate levels and are still captured without biasing the fit toward only highly active systems.

Figure~\ref{fig:close_111} examines a loading orientation close to $[111]$.
The two slip systems with the highest Schmid factors, A3 ($S=0.34$) and A6 ($S=0.32$), are much more active than all the others.
Surprisingly, D1 ($S=0.28$) and D6 ($S=0.25$) are far less active, even though their Schmid factors are not much lower.
Because A3 and A6 lie on the same slip plane, whereas D1 and D6 lie on a different one, this suggests that an active slip system tends to suppress activities of slip systems on other planes.
As in Figures~\ref{fig:close_001} and~\ref{fig:close_011}, the peak of $\dot{\gamma}_i(l)$ again lies at intermediate link length.
It is located at $l/\bar{l}_i=3.5$ for A3, and at $2.5$ for both A6 and D1.
The prediction $\dot{\gamma}_i(l)=n_i(l)\,v_i(l)\,b\,l$ again agrees very well with the DDD data.

Figure~\ref{fig:close_123} shows a loading orientation close to $[123]$.
This is a single-slip orientation.
Slip system A3 ($S=0.48$) has the highest Schmid factor of all twelve systems and is by far the most active.
The systems with the next-largest Schmid factors, A6 ($S=0.32$), B4 ($S=0.38$) and D1 ($S=0.32$), are nevertheless inactive, even though their Schmid factors are comparable to those of the systems that are active in Figure~\ref{fig:close_111}.
This suggests that a slip system (A3) that becomes too dominant suppresses all the others, including those (A6) on its own plane.
For A3, the peak of $\dot{\gamma}_i(l)$ lies at $l/\bar{l}_i=3.5$.
Across Figures~\ref{fig:close_001}--\ref{fig:close_123}, peak contributions consistently occur at intermediate normalized lengths, between about $l/\bar{l}_i=2.4$ and $3.5$.
This peak location follows directly from the competition in Eq.~\eqref{eq:Orowan_L}: short links are abundant but less effective, long links are fast but sparse, and intermediate lengths dominate total shear production.
The close match between the length-dependent slip rates $\dot{\gamma}_i(l)$ predicted from the velocity function and the DDD data is a key result of this study, confirming the proposed functional form of $v_i(l)$ in Eq.~\eqref{eq:ViL_fit}.
%
%For this near-$[123]$ case, one slip system is markedly dominant, and the fitted framework remains accurate for that dominant channel while preserving low-rate behavior on weak systems.

\subsection{Dislocation link length-based flow rule}
\label{sec:flow_rule}

\begin{figure}[htbp]
    \centering
    \edef\eqtenfigh{\the\dimexpr 0.28\textwidth\relax}
    \subcaptionbox{}{%
        \includegraphics[height=\eqtenfigh]{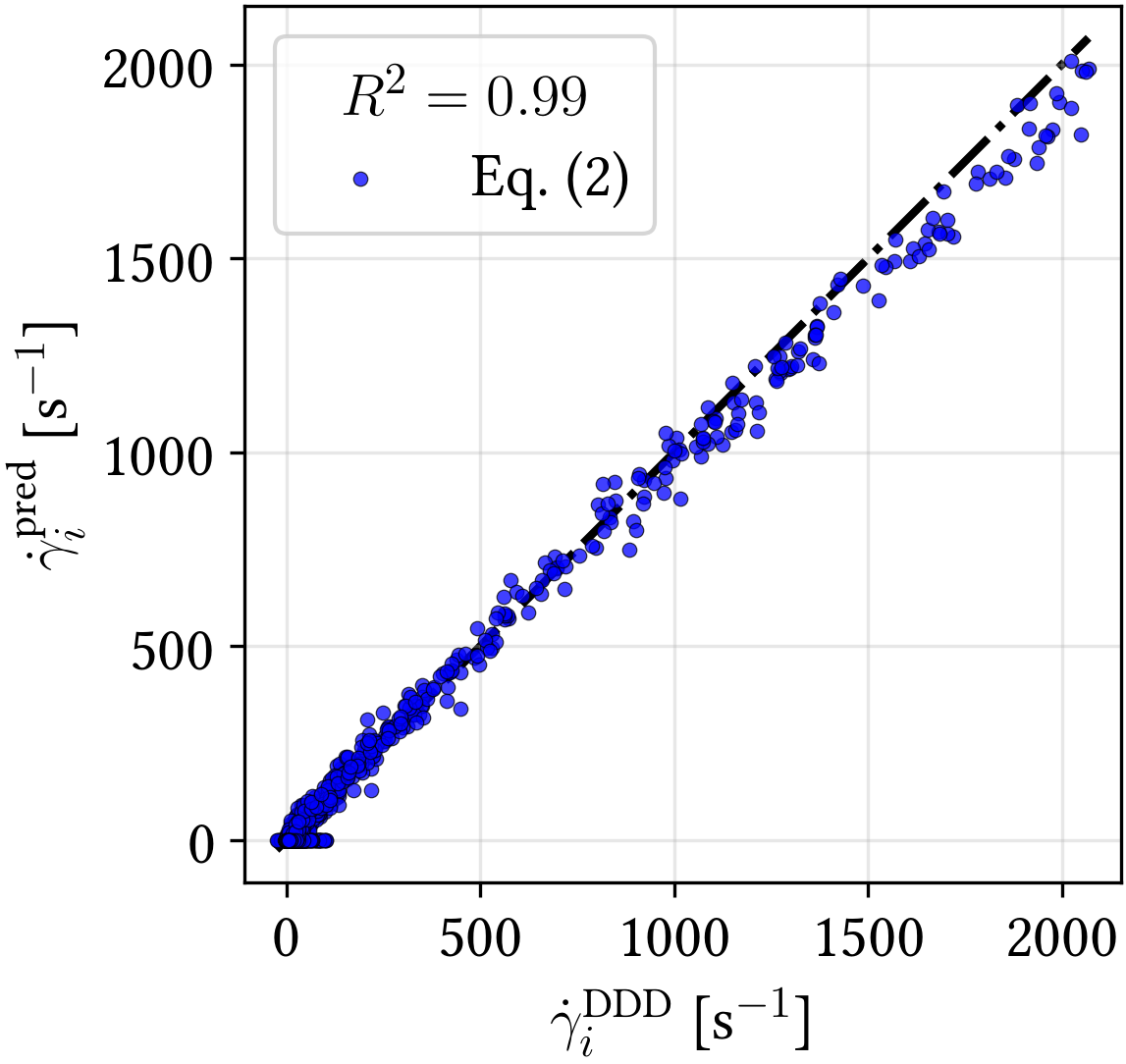}}
    \hfill
    \subcaptionbox{}{%
        \includegraphics[height=\eqtenfigh]{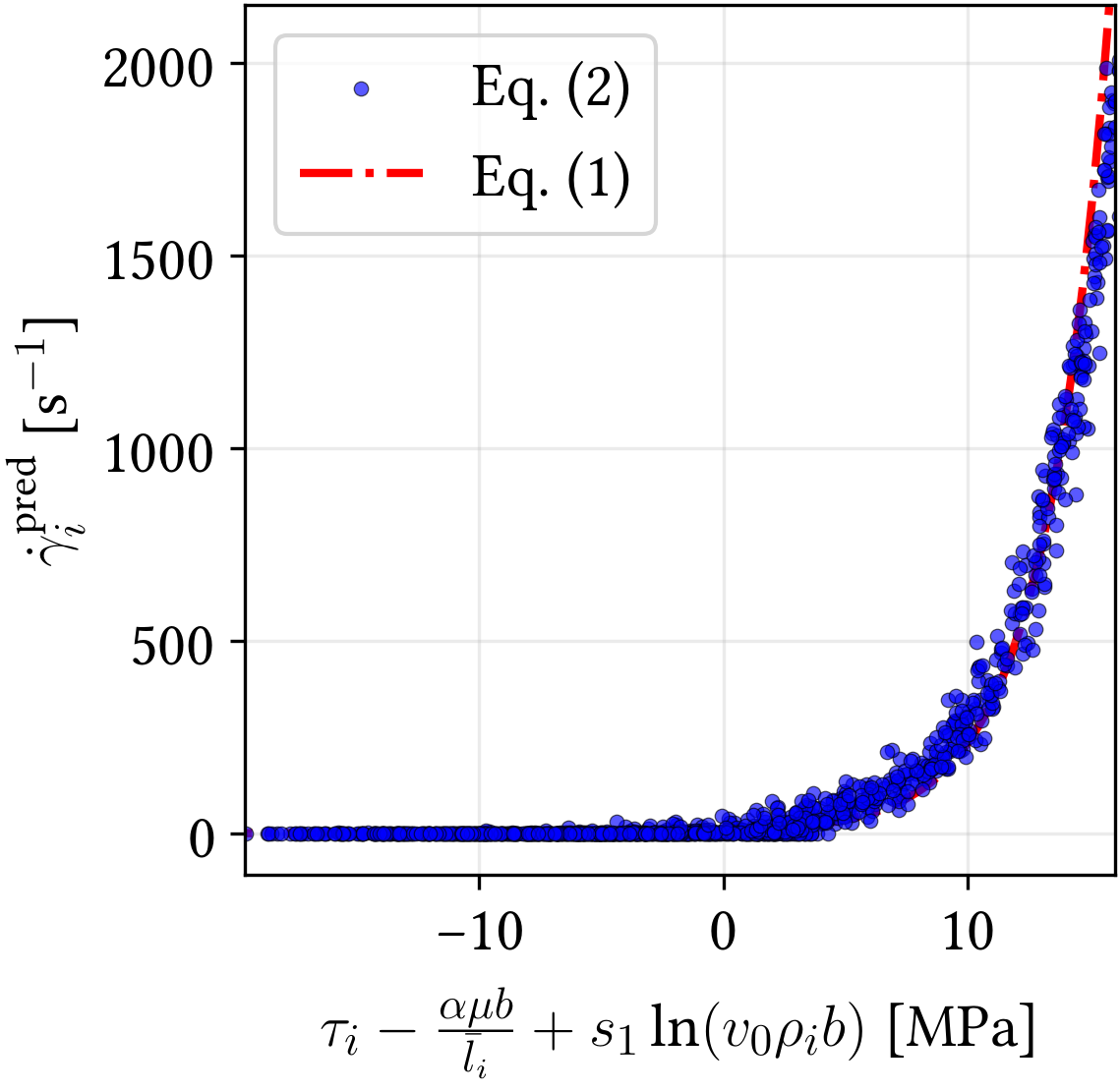}}
    \hfill
    \subcaptionbox{}{%
        \includegraphics[height=\eqtenfigh]{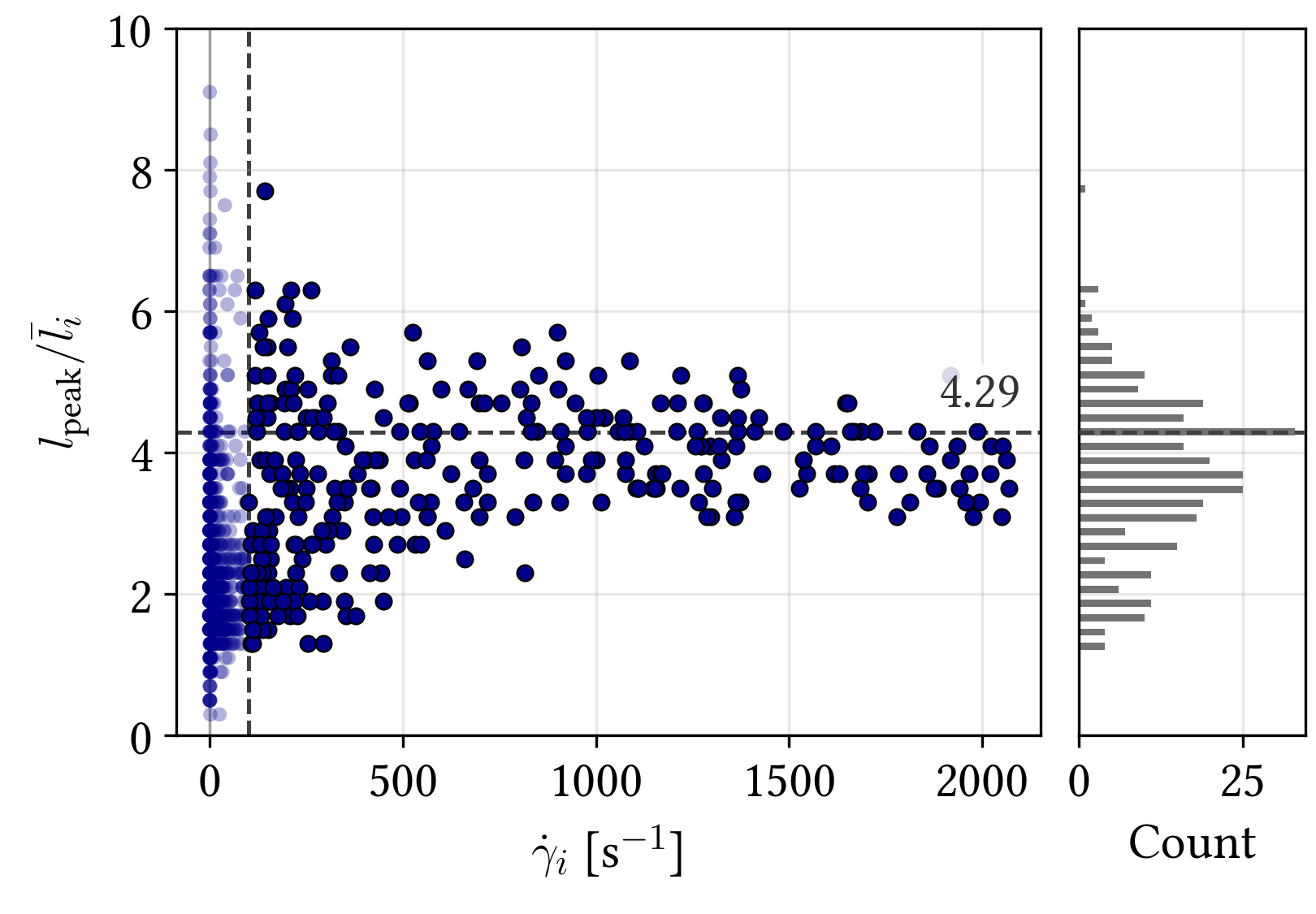}}
    \caption{%Comparison of flow-rule predictions against DDD-derived references.
    (a) Parity plot comparison of shear rate between DDD data and the model predictions based on Eqs.~\eqref{eq:Orowan_L} and \eqref{eq:ViL_fit}.
    (b) Comparison of predicted shear rate against the empirical relation, Eq.~\eqref{eq:gamma_dot_exp}.
    %with the data undergoing $\tau_i\in[8,\,9]$ MPa highlighted in orange.
    %
    (c) Normalized link length $l_{\mathrm{peak}}/\bar{l}_i$ at which $\dot{\gamma}_i(l)$ is largest, versus $\dot{\gamma}_i$, for every slip system.
    The dashed vertical line is $\dot{\gamma}_i=10^{2}\,\mathrm{s}^{-1}$, above which the slip system is considered highly active and the corresponding symbols are darker.
    %
    %Dark-blue circles with black edges lie at or above that cut.
    %
    %Faded points lie below it.
    %
    The histogram on the right corresponds to the high-rate subset, with mode $l_{\mathrm{peak}}/\bar{l}_i=4.29$.
    }
    \label{fig:gammaidot_pred}
\end{figure}

Figure~\ref{fig:gammaidot_pred}(a) is a parity plot comparing the predicted and the DDD shear rates for all $12$ slip systems of all $99$ loading orientations.
The points follow the diagonal closely, with $R^2=0.99$.
This shows that the quality of the fit over the entire dataset is very good.
Figure~\ref{fig:gammaidot_pred}(b) compares the predicted shear rates with the empirical exponential equation, Eq.~\eqref{eq:gamma_dot_exp}, as functions of the effective stress.
It is very similar to Figure~\ref{fig:gammaidot_pred_intro}, which plots the DDD shear rates against the same effective stress.
Together, Figures~\ref{fig:gammaidot_pred}(a) and~(b) justify the functional form of the velocity function in Eq.~\eqref{eq:ViL_fit} and the fitting procedure based on the loss function of Eq.~\eqref{eq:L_velocity_fit}.

%\hl{(Hanfeng has audited slip systems whose integrated $\dot{\gamma}_i$ comes out negative: 122 cases in all, listed in main\_zhai.tex.  Some have zero Schmid factor, and the rest are weakly loaded.  We need to decide whether to discuss these in the paper, and whether they should be excluded from the parity plot in (a).)}

Figure~\ref{fig:gammaidot_pred}(c) plots the peak location $l_{\rm peak}/\bar{l}_i$ of $\dot{\gamma}_i(l)$ against $\dot{\gamma}_i$, for every slip system of all $99$ loading orientations.
For the high-rate slip systems, $\dot{\gamma}_i\ge10^{2}~{\rm s}^{-1}$, the histogram of $l_{\rm peak}/\bar{l}_i$ has its mode at $4.3$, and about $80\%$ of them lie between $2$ and $5$.
The four orientations of Figures~\ref{fig:close_001}--\ref{fig:close_123} fall in the lower half of this distribution.
The spread of $l_{\rm peak}/\bar{l}_i$ narrows as $\dot{\gamma}_i$ increases.
Near $100~{\rm s}^{-1}$ it ranges from about $1.3$ to $6.3$, whereas above $1000~{\rm s}^{-1}$ it stays between about $3$ and $5.5$, close to the mode.
The peak of $\dot{\gamma}_i(l)$ thus occurs at a link length of a few times $\bar{l}_i$ on nearly every active slip system.
%
%\hl{(Check plot of fitting parameters across all loading orientations.)}

\subsection{Velocity function}
\label{sec:velocity_function}

\begin{figure}[htb]
    \centering
    \edef\eqsubfigh{\the\dimexpr 0.42\textwidth\relax}
    \begin{subfigure}[t]{0.46\textwidth}
        \centering
        \includegraphics[height=\eqsubfigh]{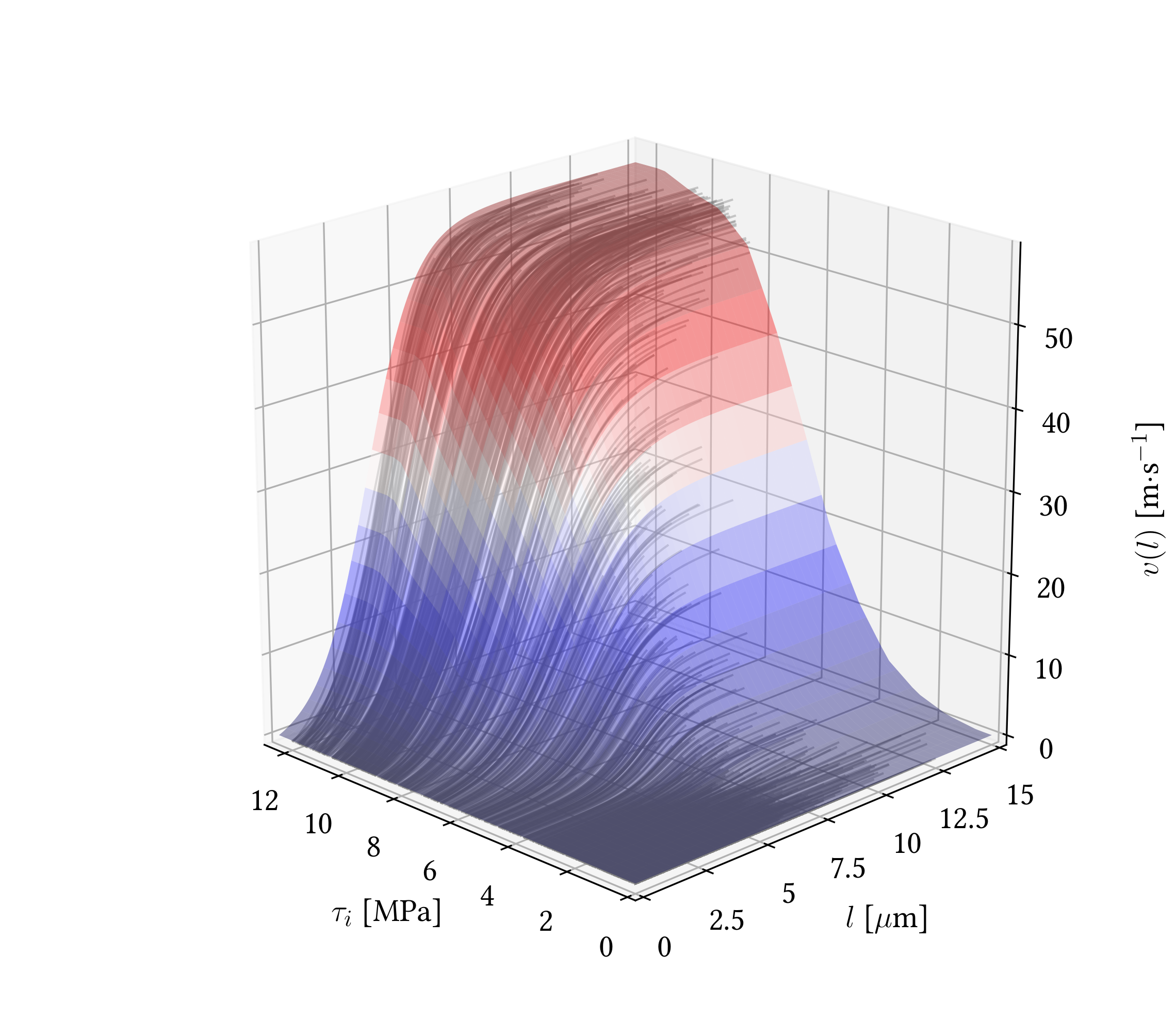}
        \caption{}
    \end{subfigure}
    \hfill
    \begin{subfigure}[t]{0.46\textwidth}
        \centering
        \includegraphics[height=\eqsubfigh]{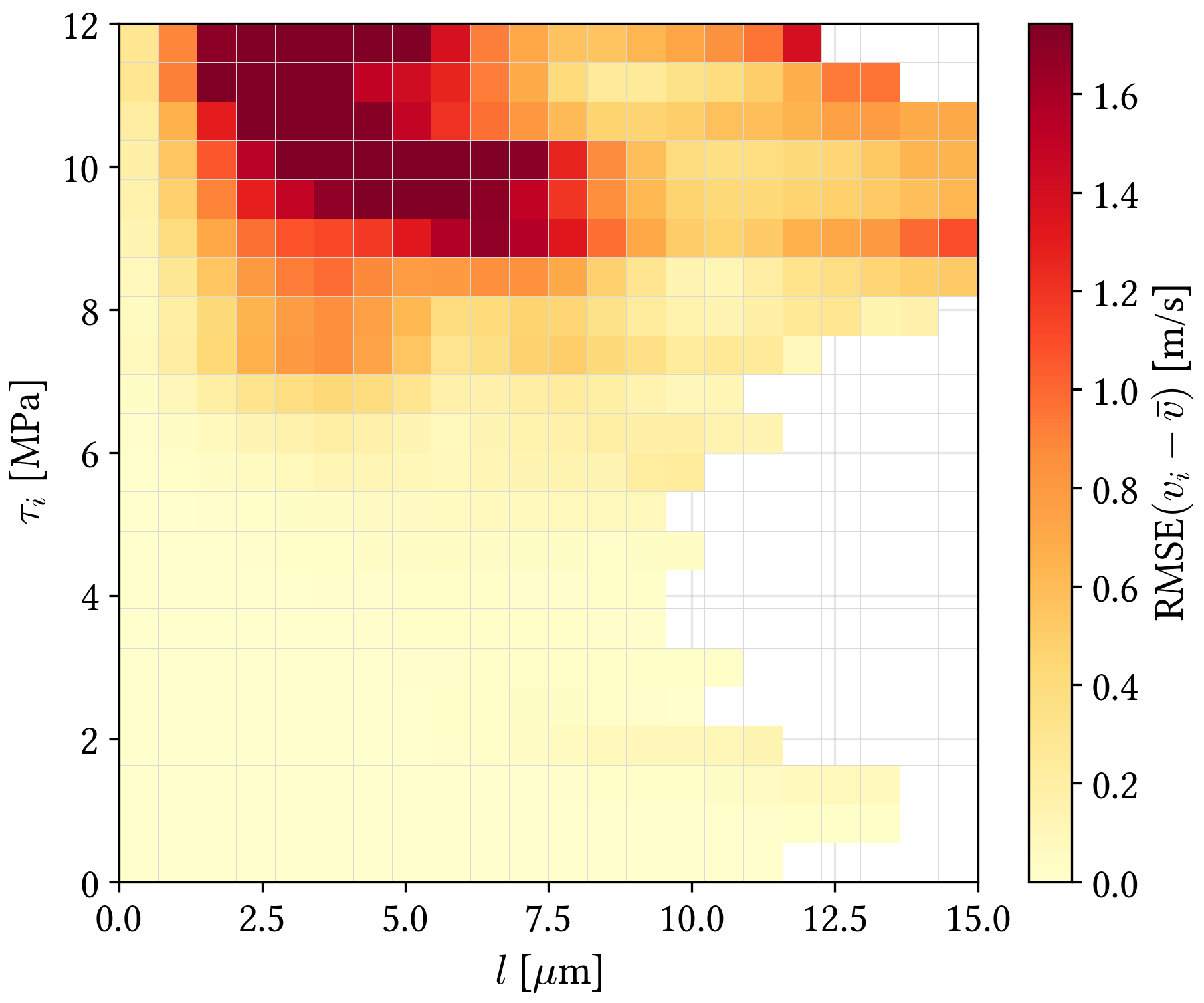}
        \caption{}
    \end{subfigure}
    \caption{(a) 
    %\hl{(Replot from a different viewing angle -- origin at bottom.)} 
    Individual velocity functions $v_i(l)$ (gray) and the mean velocity surface $\bar{v}(\tau,l)$ of Eq.~\eqref{eq:v_universal}, obtained by minimizing Eq.~\eqref{eq:L_surf}.
    (b) Square-bin map of $\mathrm{RMSE}(v_i-\bar{v})$ over the $(l,\tau_i)$ plane.
    Each curve is sampled every $0.5~\mu$m, up to $15~\mu$m or $15\,\bar{l}_i$, whichever is shorter.
    Empty cells contain no evaluated points.
    %, and the color scale ends at the 95th percentile of the non-empty cells.
    }
    \label{fig:velocity_curves}
\end{figure}

Figure~\ref{fig:velocity_curves}(a) plots the individual velocity functions $v_i(l)$ obtained in Section~\ref{sec:velocity_fit}, one gray curve for each slip system of all $99$ loading orientations.
The spread among curves at similar $\tau_i$ is small.
The curves therefore lie close to a mean velocity surface over the $(\tau_i,l)$ plane, forming a thin layer whose thickness is quantified in Figure~\ref{fig:velocity_curves}(b).
The small spread is not an artifact of the regularization in Eq.~\eqref{eq:L_velocity_fit}, although it pulls curves at similar $\tau_i$ toward one another.
If the regularization had forced the curves together, they would not reproduce the DDD data as closely as they do in Figure~\ref{fig:gammaidot_pred}(a).
The link velocities on all slip systems can therefore be well approximated by a single mean velocity surface, a surprising result that we discuss further below and in Section~\ref{sec:discussion}.
We now describe how this mean velocity surface is extracted from the individual velocity functions.

We parameterize the mean velocity surface as
\begin{equation}
    \bar{v}(\tau,l)
    =\theta^{(3)}(\tau)\,\tau
    \left[\sigma\!\left(\frac{l}{\theta^{(2)}(\tau)}-\theta^{(1)}(\tau)\right)
    -\sigma\!\left(-\theta^{(1)}(\tau)\right)\right],
    \label{eq:v_universal}
\end{equation}
which has the form of Eq.~\eqref{eq:ViL_fit} with the per-curve parameters $\theta_i^{(k)}$ replaced by shared functions $\theta^{(k)}(\tau)$, $k=1,2,3$.
Each $\theta^{(k)}(\tau)$ is a cubic Hermite spline in $\tau$ with $24$ pieces on $[0,12]$~MPa, continuous together with its first derivative.
%
%\hl{(Hanfeng to confirm.)}
%
%\hl{(The earlier text said only that the piecewise polynomial is continuous with a continuous first derivative, which is weaker than a standard cubic spline.)}
%
%\hl{(I think you mean a spline.  Need to provide number of pieces for the spline and order.  Cubic splines are most popular.)}
%
The factor $\tau$ in Eq.~\eqref{eq:v_universal} ensures $\bar{v}(0,l)=0$ for every $l$.
%
%\hl{(Hanfeng: is $\theta^{(3)}(0)=0$ still imposed?  With the explicit factor $\tau$ it is no longer needed for $\bar{v}(l;0)=0$, and it would make $\bar{v}\propto\tau^{2}$ near zero stress.)}
%
The functions $\theta^{(k)}(\tau)$ are fitted to the individual velocity functions by minimizing
\begin{equation}
    \mathcal{L}_{\mathrm{surf}}
    =\frac{1}{N}\sum_{i}
    \left\langle
    \Big[\bar{v}(\tau_i,l)-v_i(l)\Big]^{2}
    \right\rangle_{l\in\mathcal{G}_l},
    \label{eq:L_surf}
\end{equation}
where $N$ is the number of curves and $v_i(l)$ is Eq.~\eqref{eq:ViL_fit} evaluated with the per-curve parameters $\bm{\theta}_i$ held fixed.
We minimize the deviation in velocity, rather than in the parameters, because the velocity is the physical quantity that enters Eq.~\eqref{eq:Orowan_L}.
A small change in $\theta^{(k)}$ can produce a large change in $\bar{v}$, and vice versa, so a close fit in parameter space would not necessarily control the error in the velocity.

%Near vanishing resolved shear stress the velocity is nearly zero for all $l$.
%%
%The shape functions $\theta^{(1)}(\tau)$ and $\theta^{(2)}(\tau)$ are then poorly constrained by Eq.~\eqref{eq:L_surf}, and they are held at the values of the per-curve fits.
%%
%At finite stress, where $\bar{v}$ varies appreciably with $l$, the surface is controlled by Eq.~\eqref{eq:L_surf}.

The resulting mean velocity surface is drawn in Figure~\ref{fig:velocity_curves}(a) together with the individual curves.
It vanishes identically at zero stress and at zero link length, and rises monotonically with both $\tau$ and $l$.
That the link velocities of all 12 slip systems in all 99 loading orientations are well approximated by a single function of $\tau$ and $l$ is a surprising result.
It shows that, from the point of view of link velocity, the interactions between slip systems are weak.
The length-resolved shear rates $\dot{\gamma}_i(l)$, by contrast, do not collapse onto any such function of $(\tau_i,l)$.
The interactions between slip systems therefore act mainly through the link-length distribution $n_i(l)$, and through it on $\dot{\gamma}_i$.

Figure~\ref{fig:velocity_curves}(b) quantifies how closely the individual curves follow the mean velocity surface.
The difference $v_i(l)-\bar{v}(\tau_i,l)$ is sampled on a grid of $l$ with a spacing of $0.5~\mu$m.
The $(l,\tau_i)$ plane is divided into $22\times22$ cells, each about $0.68~\mu$m by $0.55$~MPa.
Each cell shows the root-mean-square of the samples from all curves falling in that cell.
Below about $7$~MPa the error stays under $0.2~\mathrm{m\cdot s^{-1}}$ everywhere.
It is largest between $9$ and $12$~MPa at link lengths of roughly $1.5$ to $8~\mu$m, where it reaches about $1.7~\mathrm{m\cdot s^{-1}}$.
Even there the error is small in relative terms, because $\bar{v}$ itself is about $50~\mathrm{m\cdot s^{-1}}$ at those stresses and link lengths.
The individual curves therefore stay within a few percent of the mean velocity surface over the whole sampled region.
The empty cells in Figure~\ref{fig:velocity_curves}(b) mark regions with no DDD data.
They lie mainly at link lengths above about $10~\mu$m for stresses below about $8$~MPa.
%, and at the longest links for the highest stresses.
%
Outside the region covered by the data, the mean velocity surface is an extrapolation, justified by the expectation that $\bar{v}$ varies smoothly with $\tau$ and $l$.

%\begin{figure}[htbp]
%    \centering
%    \includegraphics[width=0.62\linewidth]{vinf_vs_tau_universal.png}
%    \caption{The long-link limit $v_\infty$ versus $\tau_i$.
%    %
%    Each slip-system curve is a point colored by $\bar{l}_i$, with marker area proportional to $\bar{l}_i$.}
%    \label{fig:Velocity_3D}
%\end{figure}

The behavior of the mean velocity function at $l=15~\mu$m can be seen at the edge of the surface at that length in Figure~\ref{fig:velocity_curves}(a).
For such long links the self stress, which opposes the bowing out of a link between its end points, is approximately $\alpha\mu b/l\approx0.36$~MPa with $\alpha=0.39$, and is therefore negligible.
%
%\hl{(Hanfeng please double check.)}
%
%It is therefore negligible.
%
In the DDD simulations each dislocation segment follows a linear mobility law, with velocity proportional to the local stress.
The velocity at $l=15~\mu$m is nevertheless a non-linear function of the applied resolved shear stress $\tau$.
This non-linearity is caused by the internal stress from the other dislocations.
The link velocity rises appreciably only once the applied stress $\tau$ is large enough to overcome this internal stress.
That the individual curves stay close to the mean velocity surface indicates that the internal stress has something in common across all slip systems and all loading orientations.
We will explore this in the future.

\section{Discussions}
\label{sec:discussion}
%\subsection{Discussion}

% The previous, quantitative version of this subsection (Lerch transcendent and
% step-function derivations) is archived in expon_flow_rule_derive.tex and is no
% longer included.  To bring it back, uncomment the two branches below.
%\IfFileExists{expon_flow_rule_derive.tex}
%  {\input{expon_flow_rule_derive}}
%  {\input{expon_flow_rule_derive}}

\subsection{Origin of the exponential flow rule}
\label{sec:exp_origin}

Equation~\eqref{eq:Orowan_L} provides a link-based framework that explains the empirical flow rule of Eq.~\eqref{eq:gamma_dot_exp}.
Within this framework, the link velocity on every slip system is well approximated by the two-dimensional function $\bar{v}(\tau_i,l)$ of Eq.~\eqref{eq:v_universal}.
Once supplied with the link-length distributions $n_i(l)$, Eq.~\eqref{eq:Orowan_L} predicts slip rates that agree well with the DDD data, as shown in Figure~\ref{fig:gammaidot_pred}(a).
These predictions in turn reproduce the empirical relation of Eq.~\eqref{eq:gamma_dot_exp}, as shown in Figure~\ref{fig:gammaidot_pred}(b).

The two levels of description nevertheless differ in how closely the data follow them.
In Figure~\ref{fig:gammaidot_pred_intro}, the DDD slip rates still deviate appreciably from Eq.~\eqref{eq:gamma_dot_exp}, especially where the horizontal-axis value lies between $0$ and $10$~MPa.
In other words, a slip system with a large $\tau_i$ and a small $\bar{l}_i$ and one with a small $\tau_i$ and a large $\bar{l}_i$ produce similar, but not quite the same, shear rates when their values of $\tau_i-\alpha\mu b/\bar{l}_i$ are equal.
The individual link velocities deviate far less from the mean velocity surface, in relative terms, as shown in Figure~\ref{fig:velocity_curves}(b).
This indicates that the underlying physics lives on a two-dimensional manifold, the link velocity surface $\bar{v}(\tau_i,l)$.
Integrating this surface against $n_i(l)$ through Eq.~\eqref{eq:Orowan_L} reduces it to an approximately one-dimensional relation, Eq.~\eqref{eq:gamma_dot_exp}, between the shear rate per unit dislocation density, $\dot{\gamma}_i/\rho_i$, and the effective stress $\tau_i-\alpha\mu b/\bar{l}_i$.
But that reduction is only approximate, and the deviations remain significant.

\subsection{Slip systems at nearly equal resolved shear stress}
\label{sec:narrow_window}

\begin{figure}[htbp]
    \centering
    \includegraphics[width=0.40\linewidth]{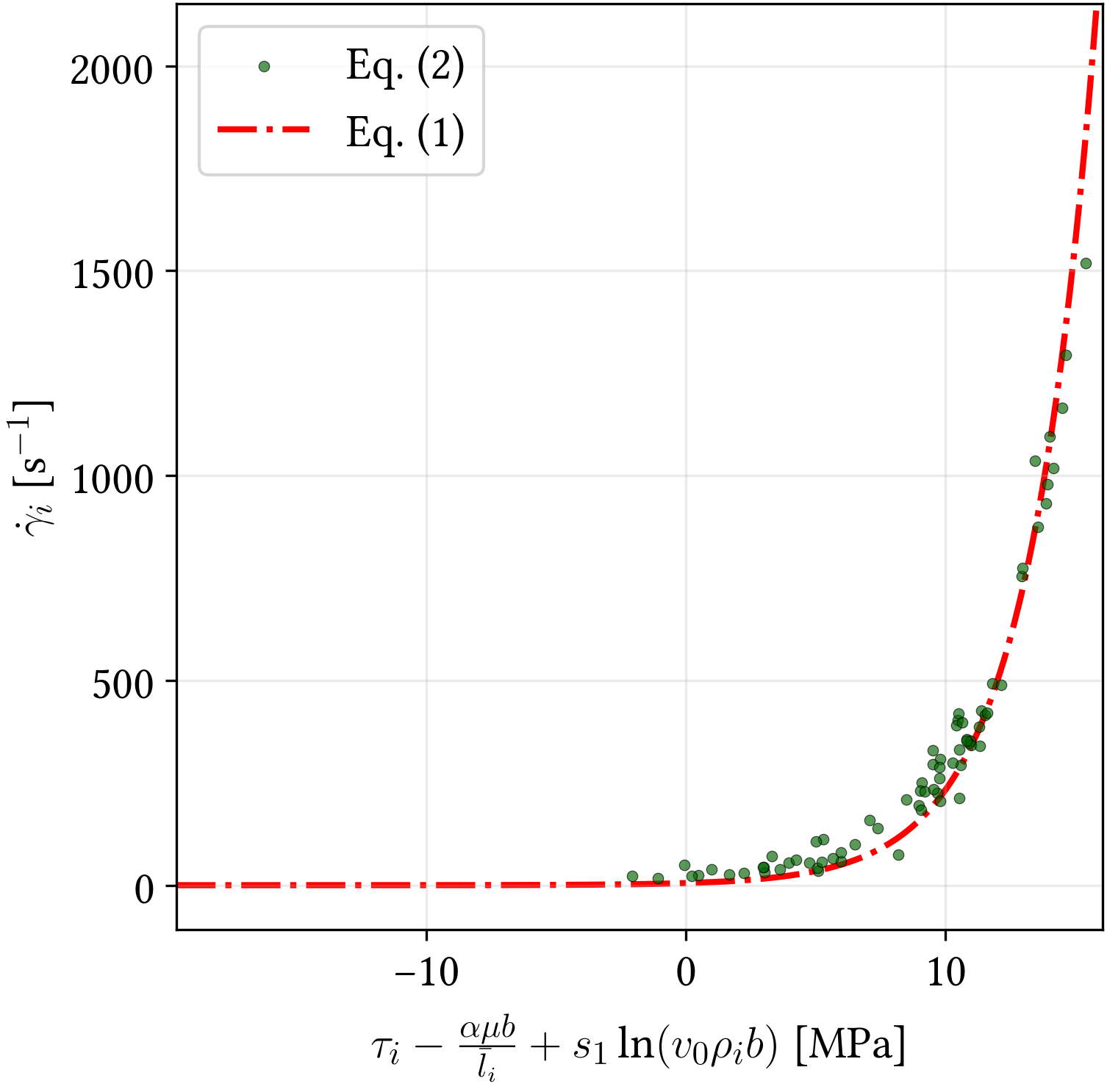}
    \caption{Predicted shear rates $\dot{\gamma}_i^{\mathrm{pred}}$ of the $73$ slip systems with $8\le\tau_i<9$~MPa.
    The points are Eq.~\eqref{eq:Orowan_L}, and the dash-dot curve is Eq.~\eqref{eq:gamma_dot_exp}.
    The horizontal axis is the same as in Figures~\ref{fig:gammaidot_pred_intro} and~\ref{fig:gammaidot_pred}(b).}
    \label{fig:narrow_window}
\end{figure}

The empirically observed exponential dependence in Eq.~\eqref{eq:gamma_dot_exp}, discussed in Section~\ref{sec:exp_origin}, means that a small change in the effective stress, $\tau_i-\alpha\mu b/\bar{l}_i$, say, by $\pm5$~MPa, can change the shear rate by orders of magnitude.
Such a change is sufficient to take a slip system from inactive to active.
Intuitively, one might expect the resolved shear stress $\tau_i$ to dominate the effective stress, with the self stress $\alpha\mu b/\bar{l}_i$ as a minor correction.
Here we show that this is not the case.
Across all slip systems and loading orientations, $\bar{l}_i$ scatters widely.
The self stress then varies over a range large enough that, at the same $\tau_i$, it alone can activate or suppress a slip system.
Figure~\ref{fig:narrow_window} illustrates this situation by limiting attention to slip systems whose resolved shear stress lies within a narrow window, $8\le\tau_i<9$~MPa.
There are $73$ such slip systems, drawn from many loading orientations.
Among these slip systems, the shear rate varies from roughly $15$ to $1500~{\rm s}^{-1}$, that is, from essentially inactive to highly active.

We now interpret this behavior through the link-based framework of Eq.~\eqref{eq:Orowan_L}.
Within this window the link velocity is essentially the same on every slip system, because $v_i(l)\approx\bar{v}(\tau_i,l)$ (Section~\ref{sec:velocity_function}).
Any difference in their shear rates must therefore come from $n_i(l)$.
Their velocity functions are indeed nearly indistinguishable.
Replacing each $v_i(l)$ by the average over the $73$ curves leaves the shear rates predicted by Eq.~\eqref{eq:Orowan_L} also in reasonably good agreement (not shown) with the DDD data.
A single velocity function thus reproduces the full transition from inactive to active behavior within the window.
This confirms that, among these slip systems, the spread in $\dot{\gamma}_i$ comes predominantly from $n_i(l)$.
Because $n_i(l)$ decays exponentially with $l$, a modest change in its tail produces a large change in the number of long links available to sweep area.
Two slip systems under the same resolved shear stress can thus differ by two orders of magnitude in shear rate simply because one has a longer tail than the other.
The interactions between slip systems barely alter the velocity of a link of given length.
Instead, they shape how many long links each slip system has, which Eq.~\eqref{eq:gamma_dot_exp} effectively captures through the self stress term $\alpha\mu b/\bar{l}_i$.
Predicting which slip systems become active therefore requires understanding how the competition among slip systems determines $n_i(l)$, a question we will pursue in the future.

\section{Conclusion}
\label{sec:conclusion}

This study establishes a slip-system flow-rule closure that links dislocation-network geometry to plastic flow through explicit link-length statistics and a stress-and-length-dependent velocity representation.
The dependence of the link velocity on link length is well described by a sigmoid function.
The link velocity also depends non-linearly on the resolved shear stress of the slip system.
Across loading orientations, the framework closely reproduces both the length-resolved and the integrated shear rates.
It also captures the large spread in shear rate between highly active and weakly active slip systems.
A robust physical mechanism emerges.
Plastic shear production is dominated by intermediate link lengths, with the peak contribution typically at a few times $\bar{l}_i$, because link abundance and link velocity compete.
%
%Compared with the previously reported effective exponential flow rule, the present closure maintains predictive quality while adding direct mechanistic interpretation and improved representation of cross-system variance.
%
These results provide support for a link-length-based approach to crystal-plasticity modeling informed by DDD simulations.
They also motivate extending this approach to other strain rates, temperatures, and initial microstructures.

Whether the details of the dislocation microstructure are necessary to predict macroscopic slip rates has been a subject of ongoing debate within the crystal plasticity community.
Our findings provide a clear answer to this question in the affirmative.
Crucially, our results suggest that the effect of the resolved shear stress on a slip system may be larger than its direct effect on the link velocity $v(\tau_i,l)$.
The collective shear stresses on all slip systems also dynamically shape the link-length distributions $n_i(l)$, which have a far greater effect on the shear rate $\dot{\gamma}_i$ than $\tau_i$ has through its direct effect on $v(\tau_i,l)$.
%
%Their long tails are filtered by the threshold-like behavior of the velocity function $v_i(l)$.
%
Consequently, our work suggests that understanding how the competition between slip systems determines the link-length distributions should be a major focus of future investigations.

\section*{Conflict of Interest}
The authors declare no competing interests.

\section*{Acknowledgment}

This work was supported by the U.S. Department of Energy, Office of Basic Energy Sciences, Division of Materials Sciences and Engineering under Award No.~DE-SC0010412 (Sh.~A. and W.~C.).
N.B.'s work was performed under the auspices of the U.S. Department of Energy by Lawrence Livermore National Laboratory under Contract DE-AC52-07NA27344.

% \section*{Data availability}

\section*{Code availability}

%\hl{(Write a concise sentence providing the URL.)}

Code and instructions for reproducing the results of this paper are available at \url{https://gitlab.com/micronano_public/Dislocation_Links}. 
% Flow_Rule_2027
Raw data will be made available on request.

\begin{comment}
The code that replays the DDD operation records and extracts the slip-system shear rates $\dot{\gamma}_i$ and the link-length data for all loading directions is available at \url{https://gitlab.com/micronano/MLmat/-/tree/disl_dynamics/DislocationDynamics/SlipActivity/LinkLengthStats}.
%
The oprec player is \texttt{stats\_\allowbreak oprec\_\allowbreak player\_\allowbreak on\_\allowbreak dict\_\allowbreak with\_\allowbreak analysis.py}, and \texttt{stats\_\allowbreak oprec\_\allowbreak player\_\allowbreak all\_\allowbreak loadings.py} runs it on every loading direction.
%
The played data are time-averaged with \texttt{stats\_\allowbreak avg\_\allowbreak link\_\allowbreak hist\_\allowbreak data.py} and collected into the fit-ready data set with \texttt{FlowRule/\allowbreak src/\allowbreak datagen/\allowbreak process\_\allowbreak gammai\_\allowbreak data.py}.
%
The \texttt{README.md} in that folder lists the commands for these steps.
%
The code that fits the flow rule and generates all figures in this paper is available at \url{https://gitlab.com/micronano/MLmat/-/tree/disl_dynamics/DislocationDynamics/SlipActivity/FlowRule}.
%
All figures are regenerated with \texttt{scripts/\allowbreak reproduce\_\allowbreak fig\_\allowbreak mc3\_\allowbreak sept2026.sh}, and the accompanying \texttt{README} lists the inputs and the figure produced by each step.
\end{comment}

\appendix

\section{Slip systems of the FCC crystal}
\label{app:slip_systems}

The 12 $\{111\}\langle110\rangle$ slip systems of the FCC crystal, labeled in Schmid--Boas notation and used throughout the main text, are listed in Table~\ref{tab:slip_systems}.
The plane and Burgers vectors are those of the DDD analysis, so the $S_i$ printed on the figures can be recomputed from this table.

\begin{table}[htbp]
    \centering
    \caption{Definition of the 12 $\{111\}\langle110\rangle$ slip systems of an FCC crystal in Schmid--Boas (SB) notation.
    Each system $i$ is specified by its slip-plane normal $\mathbf{n}_i$ and slip direction (Burgers vector) $\mathbf{b}_i=\tfrac{1}{2}\langle110\rangle$.
    These $(\mathbf{n}_i,\mathbf{b}_i)$ pairs are the DDD convention used to label the figures.}
    \label{tab:slip_systems}
    \begin{tabular}{cccc}
    \toprule
    Index $i$ & Plane normal $\mathbf{n}_i$ & Slip direction $\mathbf{b}_i$ & SB index \\
    \midrule
    1  & $(\bar{1}11)$     & $\tfrac{1}{2}[0\bar{1}1]$ & A2 \\
    2  & $(\bar{1}11)$     & $\tfrac{1}{2}[101]$       & A3 \\
    3  & $(\bar{1}11)$     & $\tfrac{1}{2}[110]$       & A6 \\
    4  & $(111)$           & $\tfrac{1}{2}[0\bar{1}1]$ & B2 \\
    5  & $(111)$           & $\tfrac{1}{2}[\bar{1}01]$ & B4 \\
    6  & $(111)$           & $\tfrac{1}{2}[\bar{1}10]$ & B5 \\
    7  & $(\bar{1}\bar{1}1)$ & $\tfrac{1}{2}[011]$       & C1 \\
    8  & $(\bar{1}\bar{1}1)$ & $\tfrac{1}{2}[101]$       & C3 \\
    9  & $(\bar{1}\bar{1}1)$ & $\tfrac{1}{2}[\bar{1}10]$ & C5 \\
    10 & $(1\bar{1}1)$     & $\tfrac{1}{2}[011]$       & D1 \\
    11 & $(1\bar{1}1)$     & $\tfrac{1}{2}[\bar{1}01]$ & D4 \\
    12 & $(1\bar{1}1)$     & $\tfrac{1}{2}[110]$       & D6 \\
    \bottomrule
    \end{tabular}
\end{table}

\section{Computation of plastic strain contribution from dislocation link}
\label{app:plastic_strain}

This appendix details how the slip-system shear rates $\dot{\gamma}_i$ and their length-resolved counterparts $\dot{\gamma}_i(l)$ are extracted from the moving dislocation network.
The extraction tracks the area swept by each link during a DDD time step~\cite{Arsenlis2007_msmse}, and supplies the quantities used in Section~\ref{sec:methods}.

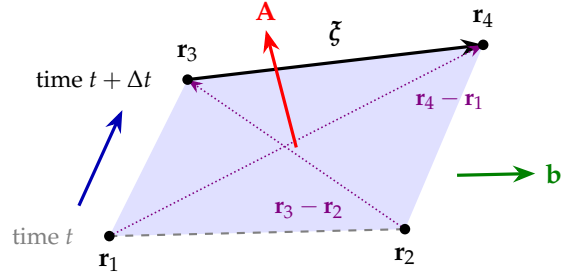
\begin{figure}[htbp]
    \centering
    \begin{tikzpicture}[>=Stealth, scale=1.15,
        dot/.style={circle, fill=black, inner sep=1.4pt}]
        % corner coordinates of the swept quadrilateral
        \coordinate (r1) at (0,0);
        \coordinate (r2) at (3.4,0.08);
        \coordinate (r3) at (0.9,1.8);
        \coordinate (r4) at (4.3,2.2);
        % swept area (shaded)
        \fill[blue!12] (r1) -- (r2) -- (r4) -- (r3) -- cycle;
        % segment at time t
        \draw[gray, dashed, thick] (r1) -- (r2);
        % segment at time t + dt, with line direction xi
        \draw[black, very thick, ->] (r3) -- (r4) node[midway, above=2pt] {$\bm{\xi}$};
        % diagonals = cross-product factors
        \draw[densely dotted, ->, semithick, violet] (r1) -- (r4)
            node[pos=0.80, below right=-1pt] {\small $\mathbf{r}_4-\mathbf{r}_1$};
        \draw[densely dotted, ->, semithick, violet] (r2) -- (r3)
            node[pos=0.25, below left=-1pt] {\small $\mathbf{r}_3-\mathbf{r}_2$};
        % endpoint nodes
        \node[dot, label={[label distance=1pt]below:{$\mathbf{r}_1$}}] at (r1) {};
        \node[dot, label={[label distance=1pt]below:{$\mathbf{r}_2$}}] at (r2) {};
        \node[dot, label={[label distance=1pt]above:{$\mathbf{r}_3$}}] at (r3) {};
        \node[dot, label={[label distance=1pt]above:{$\mathbf{r}_4$}}] at (r4) {};
        % swept-area vector A (out of the swept surface)
        \coordinate (c) at ($0.25*(r1)+0.25*(r2)+0.25*(r3)+0.25*(r4)$);
        \draw[red, very thick, ->] (c) -- ++(-0.35,1.35) node[above] {$\mathbf{A}$};
        % Burgers vector
        \draw[green!50!black, very thick, ->] (4.0,0.7) -- ++(0.9,0.02) node[right] {$\mathbf{b}$};
        % time annotations
        \node[gray, anchor=east] at ($(r1)+(-0.3,0)$) {\small time $t$};
        \node[anchor=east] at ($(r3)+(-0.3,0)$) {\small time $t+\Delta t$};
        % sweep direction (left)
        \draw[blue!70!black, very thick, ->] ($(r1)+(-0.35,0.35)$) -- ($(r3)+(-0.75,-0.35)$);
    \end{tikzpicture}
    \caption{Schematic of the swept-area construction for one moving dislocation segment.
    The segment is $\mathbf{r}_1\!\to\!\mathbf{r}_2$ at time $t$ and $\mathbf{r}_3\!\to\!\mathbf{r}_4$ at time $t+\Delta t$ (line direction $\bm{\xi}$), sweeping the shaded quadrilateral.
    The swept-area vector $\mathbf{A}$ is normal to this surface and is constructed from the two quadrilateral diagonals (Eq.~\eqref{eq:swept_area_vector}).
    The green arrow marks the Burgers' vector $\mathbf{b}$.
    The blue arrow marks the sweeping direction.
    Reversing the line direction relabels the endpoints and changes both $\mathbf{A}$ and $\mathbf{b}$ by a sign, leaving $\mathbf{A}\otimes\mathbf{b}+\mathbf{b}\otimes\mathbf{A}$ unchanged.}
    \label{fig:swept_area_schematic}
\end{figure}

The network is represented as a graph of nodes connected by straight segments; each segment carries a Burgers vector $\mathbf{b}$.
For a given segment, let $\mathbf{r}_1$ and $\mathbf{r}_2$ denote the endpoint positions at the current time step $t+\Delta t$, ordered along the line direction $\bm{\xi}$.
Let $\mathbf{r}_3$ and $\mathbf{r}_4$ denote the positions of the same two endpoints at the previous time step $t$.
Periodic boundary conditions are applied when forming endpoint differences.
The quadrilateral swept by the moving segment defines a swept-area vector
\begin{equation}
    \mathbf{A}=\tfrac{1}{2}(\mathbf{r}_3-\mathbf{r}_2)\times(\mathbf{r}_4-\mathbf{r}_1),
    \label{eq:swept_area_vector}
\end{equation}
whose magnitude $\|\mathbf{A}\|$ is the scalar swept area of that segment, as illustrated in Figure~\ref{fig:swept_area_schematic}.
Summing the symmetric dyadic products over all segments $q$ in the dislocation network and normalizing by the volume $V$ gives the incremental plastic strain tensor:
\begin{equation}
    \Delta\varepsilon^{\mathrm{pl}}
    =
    \frac{1}{2V}\sum_{q}
    \big(\mathbf{A}\otimes\mathbf{b}+\mathbf{b}\otimes\mathbf{A}\big)_q\quad,
    \label{eq:delta_eps_pl_sweep}
\end{equation}
where $\otimes$ denotes the dyadic (outer) product.
This construction ensures $\Delta\varepsilon^{\mathrm{pl}}$ is symmetric.
The line direction $\bm{\xi}$ enters Eq.~\eqref{eq:swept_area_vector} through the endpoint ordering.
Therefore, reversing both $\bm{\xi}$ and $\mathbf{b}$, which gives an equivalent description of the same dislocation, changes the signs of both $\mathbf{A}$ and $\mathbf{b}$ but leaves Eq.~\eqref{eq:delta_eps_pl_sweep} unchanged (Figure~\ref{fig:swept_area_schematic}).

The segment contributions to $\Delta\varepsilon^{\mathrm{pl}}$ on dislocation link $k$ are summed to form $\Delta\varepsilon^{\mathrm{pl}}_{k}$.
Slip system $i$ has unit slip-plane normal $\hat{\mathbf{n}}_i=\mathbf{n}_i/\|\mathbf{n}_i\|$ and unit slip direction $\hat{\mathbf{b}}_i=\mathbf{b}_i/\|\mathbf{b}_i\|$ (Table~\ref{tab:slip_systems}).
Under uniaxial loading along the unit vector $\hat{\mathbf{e}}$, the resolved shear stress computed with the pair $(\hat{\mathbf{n}}_i,\hat{\mathbf{b}}_i)$ of Table~\ref{tab:slip_systems} is proportional to $(\hat{\mathbf{n}}_i\cdot\hat{\mathbf{e}})(\hat{\mathbf{b}}_i\cdot\hat{\mathbf{e}})$.
It can be negative for some slip systems and loading orientations.
In this paper, however, $\tau_i$ is taken to be non-negative.
For those slip systems the slip direction is reversed, which we account for through the sign $s_i=\mathrm{sgn}\big[(\hat{\mathbf{n}}_i\cdot\hat{\mathbf{e}})(\hat{\mathbf{b}}_i\cdot\hat{\mathbf{e}})\big]$.
For a link $k$ belonging to slip system $i$, the incremental plastic shear strain is obtained by projecting $\Delta\varepsilon^{\mathrm{pl}}_{k}$ onto the pair $(\hat{\mathbf{n}}_i,s_i\hat{\mathbf{b}}_i)$:
\begin{equation}
    \Delta\gamma_i^{(k)} = 2\,s_i\,\hat{\mathbf{n}}_i^{\mathsf T}\,\Delta\varepsilon^{\mathrm{pl}}_{k}\,\hat{\mathbf{b}}_i,
    \label{eq:Delta_gamma_from_eps}
\end{equation}
where the factor $2$ corresponds to the engineering shear-strain convention $\gamma_{ij}=2\varepsilon_{ij}$.
Dividing by the DDD time-step interval $\Delta t$ between snapshots gives the contribution $\dot{\gamma}_i^{(k)}=\Delta\gamma_i^{(k)}/\Delta t$ of link $k$ to the shear rate.
Each link is assigned to the slip system defined by its Burgers vector and its slip plane, and is counted once.
The total shear rate on slip system $i$ is the algebraic sum of these signed contributions over all links on that system, %( with no absolute value).
\begin{equation}
    \dot{\gamma}_i = \sum_{k\in\mathcal{L}_i}\dot{\gamma}_i^{(k)},
    \label{eq:dotgamma_from_DDD}
\end{equation}
where $\mathcal{L}_i$ is the set of links on slip system~$i$.
Even with this sign convention, $\dot{\gamma}_i$ can still be negative, but its magnitude is then very small, as Figures~\ref{fig:gammaidot_pred_intro} and~\ref{fig:gammaidot_pred} show.
The link contributions $\dot{\gamma}_i^{(k)}$ are further binned by normalized link length $l/\bar{l}_i$.
\printbibliography

\end{document}